\documentclass[prb,aps,twocolumn,showpacs,floats]{revtex4}
\usepackage{graphicx}  % Include figure files
\usepackage{dcolumn}   % Align table columns on decimal point
\usepackage{amsmath}
\usepackage{amssymb}

\begin{document}

%------------------------------------------------------------------------------
\title{Composition, structure, and stability of the rutile TiO$_2$(110)
       surface: oxygen depletion, hydroxylation, hydrogen migration and
       water adsorption}

\author{Piotr M. Kowalski$^1$, Bernd Meyer$^{1,2}$ and Dominik Marx$^1$}

\affiliation{\strut$^1$Lehrstuhl f\"ur Theoretische Chemie,
             Ruhr-Universit\"at Bochum, 44780 Bochum, Germany\\
             \strut$^2$Interdisziplin\"ares Zentrum f\"ur Molekulare
             Materialien (ICMM) and Computer-Chemie-Centrum (CCC),
             Universit\"at Erlangen-N\"urnberg, 91052 Erlangen, Germany}

\date{\today}
%------------------------------------------------------------------------------

\begin{abstract}
A comprehensive phase diagram of lowest-energy structures and compositions
of the rutile TiO$_2$(110) surface in equilibrium with a surrounding gas
phase at finite temperatures and pressures has been determined using density
functional theory in combination with a thermodynamic formalism. The exchange
of oxygen, hydrogen, and water molecules with the gas phase is considered.
Particular attention is given to the convergence of all calculations with
respect to lateral system size and slab thickness. In addition, the
reliability of semilocal density functionals to describing the energetics
of the reduced surfaces is critically evaluated. 
For ambient conditions the surface is found to be fully covered by
molecularly adsorbed water. At low coverages, in the limit of single,
isolated water molecules, molecular and dissociative adsorption become
energetically degenerate.
Oxygen vacancies form in strongly reducing, oxygen-poor environments.
However, already at slightly more moderate conditions it is shown that
removing full TiO$_2$ units from the surface is thermodynamically preferred.
In agreement with recent experimental observations it is furthermore
confirmed that even under extremely hydrogen-rich environments the surface
cannot be fully hydroxylated, but only a maximum coverage with hydrogen of
about 0.6--0.7 monolayer can be reached.
Finally, calculations of migration paths strongly suggest that hydrogen
prefers to diffuse into the bulk over desorbing from the surface into the
gas phase. 
\end{abstract}

\pacs{%
68.43.Fg, %  Adsorbate structure (binding sites, geometry)
68.43.Bc, %  Ab initio calculations of adsorbate structure and reaction
68.47.Gh, %  Oxide surfaces
82.65.+r  %  Surface and interface chemistry; heterogeneous catalysis at surf.
}

\maketitle

%------------------------------------------------------------------------------
\section{Introduction}
%------------------------------------------------------------------------------

Over the past years, the (110) surface of TiO$_2$ in the rutile structure
has become one of the most popular model systems for fundamental surface
science studies of transition metal oxides.\cite{DB03}  It is the
thermodynamically most stable crystal face of TiO$_2$ and therefore
represents the dominating facet of rutile crystallites.\cite{DV94}
Stoichiometric, single crystal (1$\times$1) surfaces may be easily prepared,
and most experimental surface science techniques can be applied without
difficulties.\cite{DB03}  The interest in TiO$_2$(110) surfaces is furthermore
driven by many technological applications of TiO$_2$, ranging from pigments,
coatings, electronic devices, implants, gas sensors, photochemical reactions
to catalysis. In all of them, the surface properties of TiO$_2$ play a crucial
role.\cite{DB03}

One of the most important properties of TiO$_2$ is that it can be easily
reduced. The reducibility is essential for many applications of TiO$_2$
in heterogeneous catalysis. Oxide-supported metal-based catalysts with TiO$_2$
as part of the support often show a so-called strong metal--support interaction
(SMSI).\cite{TA78}  Here, the catalytic properties of the supported metal
clusters are profoundly modified by an incorporation of partially reduced
TiO$_x$ into the boundary areas of the metal particles.\cite{TA87}  In
many cases, the TiO$_x$-decorated metal particles exhibit a much higher
catalytic activity for hydrogenation reactions than the pure metal
itself.\cite{VS84}

One way to reduce the TiO$_2$(110) surface is to remove surface O atoms.
By this process, formally two neighboring Ti$^{4+}$ ions of a vacancy change
to a Ti$^{3+}$ oxidation state. In ultra-high vacuum (UHV) experiments
O~vacancies are easily created either by electron bombardment, sputtering,
or simply by annealing. The presence of O~vacancies strongly increases the
reactivity of the surface. Among the many investigated processes, probably 
the best studied surface reaction is the dissociation of water, which
has been shown to occur at O~vacancies,\cite{ExW1,ExW2,ExW3,ExW4,ExW5}
whereas on the stoichiometric, well-annealed parts of the surface the
water molecules stay mostly intact.\cite{ExW6,ExW7,ExW8}  There is, however,
a limit to what extend the surface can be reduced. Oxygen vacancy
concentrations are typically in the order of several percent,\cite{DB92,ExW7}
but it is not possible to remove all surface O atoms, in contrast to, for
example, the rutile SnO$_2$(101) surface.\cite{DB05,BM07}

Alternatively, the TiO$_2$(110) surface can be reduced by hydroxylation of
the surface O atoms via adsorption of hydrogen. While on unreducible oxides,
such as MgO, a heterolytic dissociative adsorption of H$_2$ with H$^+$
adsorbing on O$^{2-}$ and H$^-$ on the metal cations is preferred, it is more
favorable for reducible oxides, such as TiO$_2$, to form only OH$^-$ groups.
In the latter case the excess electrons are transferred to the cations, thus
reducing Ti$^{4+}$ to Ti$^{3+}$.

The interaction of hydrogen with TiO$_2$(110), however, has been much less
intensively investigated by surface science studies than the reduction by
O depletion. This is rather surprising in view of the importance of TiO$_2$
as catalyst component for hydrogenation reactions and the prospective
application of TiO$_2$ as photocatalyst for the decomposition of water.
It has been shown that molecular hydrogen does not interact strongly with
TiO$_2$(110),\cite{HK81,KB04}  while atomic hydrogen readily sticks to
the surface O atoms.\cite{KB04,SF00,FK01,YC08}  No Ti--H vibrations could
be detected with high resolution electron energy loss spectroscopy
(HREELS).\cite{YC08}  Interestingly, also the reducibility of TiO$_2$(110)
with hydrogen is limited. Recent scanning tunneling microscopy (STM)
measurements\cite{YC08} revealed that even after very high exposures to
atomic hydrogen the surface cannot be saturated. Only a maximum coverage
of about 0.7 monolayers could be achieved. The scattering of thermal energy
He atoms (HAS) showed that the hydroxyl groups do not form an ordered
overlayer.\cite{KB04}  Performing a thermal desorption spectroscopy (TDS)
experiment by monitoring the He atom reflectivity of the surface while
increasing the temperature (He-TDS), two distinct changes in the reflectivity
at 388\,K and 626\,K were observed, which indicate structural rearrangements
of the surface.\cite{KB04}  These rearrangements must involve the loss of the
hydrogen atoms, since afterwards the surface was H--free, as seen by HAS and
HREELS. However, in conventional TDS no desorption of H$_2$, and only a very
small amount of H$_2$O was detected,\cite{YC08} even after heating the surface
up to 650\,K. This led to the conclusion that the H atoms rather diffuse into
the bulk than desorb from the surface.\cite{YC08}

As for experiment, also theoretical studies of reduced TiO$_2$(110) have
mainly focused on O deficient surfaces. While the properties of O~vacancies
on TiO$_2$(110) have been investigated extensively\cite{Ovac0,Ovac1,Ovac2,%
Ovac3,Ovac4,Ovac5,Ovac6,Ovac7,Ovac8,Ovac9,OvacA,OvacB,OvacC} (see
Ref.~\onlinecite{GP07} for a recent review), we are aware of only two recent
studies addressing specifically the adsorption of hydrogen.\cite{YC08,LM02}
The determination of binding energies, adsorption sites and some reaction
barriers, however, was restricted in both cases to slabs with a (1$\times$1)
periodicity, thus assuming H coverages of one and more H atoms per surface
unit cell.

Finally, next to the formation of O~vacancies and the hydroxylation of
the surface by hydrogen adsorption, also the interaction with water has
to be taken into account. The adsorption of water, either molecular or
dissociative, does not lead to a reduction of the surface. But since water
is always present, even in well-controlled UHV experiments, one has to
account for residual hydroxyl groups which affect other adsorption and
reaction processes.\cite{DB03}  Water adsorption on TiO$_2$(110) has been
investigated extensively both experimentally and theoretically. Among the
experimental studies the common view is that at all coverages water adsorbs
molecularly on the ideal terraces of the TiO$_2$(110) surface and only
dissociates at defects.\cite{DB03,ExW6,ExW7,ExW8}  However, it should be
noted that in all these experiments always a small amount of dissociated water
was present as seen by XPS and HREELS and indicated by a high-temperature tail
in TDS. These signatures were naturally attributed to water molecules
dissociated at O~vacancies, but it would be difficult to distinguish them from
a situation in which water initially would adsorb dissociatively at very low
coverages and molecularly afterwards as suggested by some early studies.%
\cite{ExW6,ExW9}   In contrast to the consensus among the experimental studies
the results from theoretical calculations are very contradictorily. Early
investigations predicted water dissociation at all coverages in complete
disagreement with experiment. Only in some more recent calculations partial
dissociated or molecular structures of water were found to be lower in energy.
Overall, the theoretical studies are almost evenly divided whether molecular,
dissociative or a partial dissociative adsorption of water (strongly
depending on the coverage) is predicted as the energetically most stable
state.\cite{ThW1,ThW2,ThW3,ThW4,ThW5,ThW6,ThW7,ThW8,ThW9,ThWA}

\begin{figure}[!t]
\noindent
\includegraphics[width=200pt]{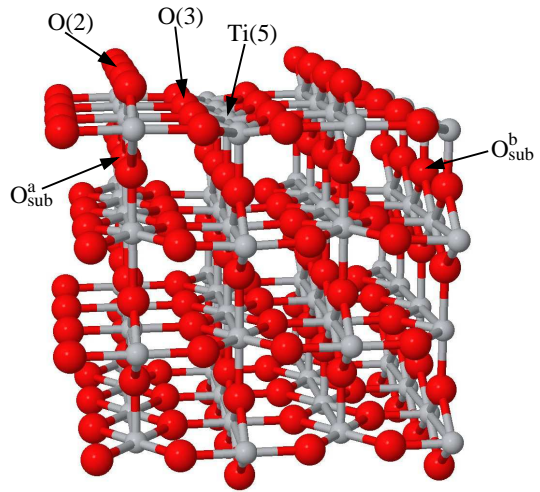}
\caption{\label{F1}
Atomic structure of the stoichiometric rutile TiO$_2$(110) surface. Oxygen
atoms are shown in red, Ti atoms in gray. The two- and threefold-coordinated
O sites and the fivefold-coordinated Ti sites are labeled by O(2), O(3) and
Ti(5), respectively.}
\end{figure}

The aim of the present paper is to explore the competition between the
reduction of the TiO$_2$(110) surface via O~vacancy formation or hydrogen
adsorption and the non-reductive interaction of the surface with water in
a comprehensive and systematic way. For a series of surface models with
different O defect configurations and for a wide range of both hydrogen and
water coverages we determine the total energies and the fully relaxed atomic
structures using first-principles density functional theory (DFT). In order
to extend the zero temperature and zero pressure DFT results to relevant
environmental situations, such as UHV or reaction conditions in heterogenous
catalysis, and to identify the thermodynamically most stable surface
structures and compositions depending on the experimental conditions, we
assume that the surfaces are in thermodynamic equilibrium with a surrounding
gas phase at a given temperature $T$ and finite partial pressures $p$.
To account for the exchange of oxygen and hydrogen between the surface
and the gas phase, appropriate chemical potentials $\mu_{\rm O}(T,p)$ and
$\mu_{\rm H}(T,p)$ are introduced.\cite{KP87,QM88,RS01,BM04}  By minimizing
the Gibbs free surface energy as function of the chemical potentials, surface
phase diagrams of the most stable surface structure and composition are
constructed depending on both temperature and partial pressures.\cite{KP87,%
QM88,RS01,BM04} 

Within this context the main focus of the present study will be on the
hydroxylated surface. In addition to the thermodynamic considerations we
investigate the kinetic behavior of hydrogen atoms and water molecules on the
TiO$_2$(110) surface. Energy barriers for various surface processes related to
the migration and desorption of hydrogen and to the dissociation of water are
calculated. Our results support the recently proposed suggestion that hydrogen
atoms, instead of being desorbed from the surface at higher temperatures,
migrate into the bulk. In addition, we find a rather small barrier for water
dissociation. However, in contrast to some previous studies, our results
indicate that the molecular adsorption of water is preferred over dissociation
in the monolayer coverage limit.

%------------------------------------------------------------------------------
\section{Computational Approach}
\label{method}
%------------------------------------------------------------------------------

The DFT calculations for the different TiO$_2$(110) surface structures, as
well as the bulk and molecular reference energies, have been carried out
using the Car--Parrinello Molecular Dynamics (CPMD) code.\cite{CP1,CP2}
The gradient-corrected Perdew--Burke--Ernzerhof functional\cite{PBE}  PBE
was used to describe the exchange and correlation effects. Vanderbilt
ultrasoft pseudopotentials\cite{DV90} were employed together with a plane
wave basis set to represent the Kohn-Sham orbitals. A plane wave cut-off
energy of 25\,Ryd was sufficient to get well converged results for structures
and energetics. In particular, increasing the cut-off to 30\,Ryd changed the
adsorption energies for water molecules by less than 0.02\,eV. The Ti
pseudopotential was constructed from an ionic 3$d^1$\,4$s^2$ configuration,
and the 3$s$ and 3$p$ semicore electrons were treated as full valence states.
Since very large supercells were used (see below), the $k$--point sampling was
restricted to the $\Gamma$--point. Spin polarization was included for all
systems with an odd number of electrons and for all calculations with oxygen
vacancies. All configurations were relaxed by minimizing the atomic forces.
Convergence was assumed when the maximum component of the residual forces on
the ions was less than 0.01\,eV/\AA. 

With this computational setup we find for the optimized TiO$_2$ bulk
lattice parameters values of $a$\,=\,4.649\,{\AA}, $c$\,=\,2.966\,{\AA}
and $u$\,=\,0.305, which compares very well to previous GGA calculations%
\cite{Ovac1,Ovac8,Ovac9,ThW2,ThW7,BG04,TL06}  and to experiment
($a$\,=\,4.594\,{\AA}, $c$\,=\,2.959\,{\AA} and $u$\,=\,0.305, see
Ref.~\onlinecite{AB71}).

\begin{figure}[!t]
\noindent
\includegraphics[width=246pt]{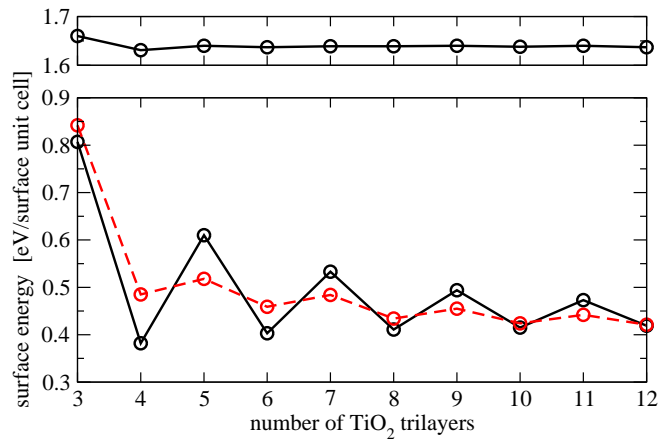}
\caption{\label{F2}
Convergence of the unrelaxed (upper panel) and relaxed (lower panel)
TiO$_2$(110) surface energy with slab thickness; note the different
energy scale. The solid black line in the lower panel represents the
result from fully relaxed slabs. The calculation with the atoms in the
bottom two trilayers fixed at the bulk positions is shown by the red
dashed line.}
\end{figure}

\begin{figure}[!t]
\noindent
\includegraphics[width=246pt]{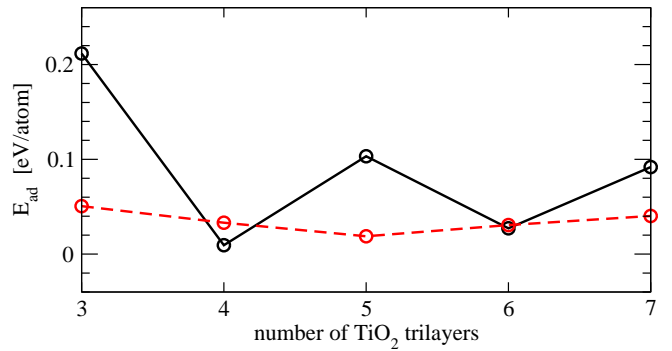}
\caption{\label{F3}
Convergence of the hydrogen adsorption energy $E_{\rm ad}$ with slab thickness
at full monolayer coverage. The atoms in the bottom two trilayers were fixed
at the bulk positions. The solid black and the dashed red lines distinguish
calculations without and with saturation of the broken surface bonds at the
bottom of the slabs with pseudo atoms of nucleus charge of +4/3 and +2/3,
respectively.}
\end{figure}

\begin{figure}[!t]
\noindent
\includegraphics[width=246pt]{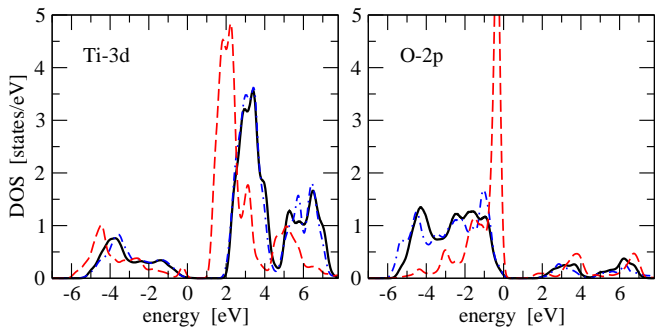}
\caption{\label{F4}
Local density of states (LDOS) of Ti and O atoms in bulk TiO$_2$ (black solid
lines), of the fivefold- and twofold-coordinated Ti(5) and O(2) surface
atoms at the bottom of a four trilayer TiO$_2$(110) slab (top two layers
relaxed, bottom two layers fixed at bulk positions, red dashed lines), and
after saturation of the bottom Ti(5) and O(2) surface atoms with
pseudo-hydrogen atoms (see text, blue dashed-dotted lines).}
\end{figure}

For the calculation of the energetics of O~vacancy formation as well as
hydrogen and water adsorption, the total energies of the isolated O$_2$,
H$_2$ and H$_2$O molecules are needed. While H$_2$ and H$_2$O are reasonably
well described within PBE/DFT, it is a well-known deficiency of all local
and semilocal functionals (such as PBE) that they strongly overbind the O$_2$
molecule.\cite{BA00,RS01,Ovac4,Ovac6,Ovac7,GP07}  In our setup we find an
O$_2$ binding energy of 5.87\,eV, which is about 0.4\,eV lower than the fully
converged PBE value of 6.24\,eV,\cite{PBE,Com1}  which is in turn about 1\,eV
larger than the experimental value\cite{NIST}  of 5.26\,eV (with zero point
vibrational energy removed in the harmonic approximation).  

In order to circumvent errors introduced by a poor description of the O$_2$
molecule, we do not employ the total energy of the O$_2$ molecule from the
DFT calculation as reference, but deduce it via a thermodynamic cycle from
the total energies of the H$_2$ and H$_2$O molecules together with the
experimental value of 2.51\,eV\cite{NIST} for the formation energy of water
from H$_2$ and O$_2$ (also taking into account the corrections due to zero
point vibrations).
With this reference energy for the gas phase O$_2$ molecule we obtain a
formation energy of bulk TiO$_2$ from metallic bulk Ti and O$_2$ of 9.60\,eV,
which is in excellent agreement with the experimental formation enthalpy of
9.73\,eV.\cite{NIST}

The second problem we have to address is how reliable can we expect the
DFT calculations to be when describing the reduced state of an
insulator/semiconductor.\cite{P08}  It is a well-known shortcoming of LDA
and all commonly used GGA functionals that band gaps are substantially
underestimated due to the insufficient cancellation of the self-interaction
energy. With our setup we find a band gap for bulk TiO$_2$ of 1.86\,eV, which
is in very good agreement with previous DFT calculations,\cite{GC92,Ovac0,%
Ovac1,Ovac2,BG04}  but is much smaller than the experimental value of
3.03\,eV,\cite{PC77}  as expected. Upon reduction, defect states are created,
which typically appear in the band gap region. In DFT calculations, however,
due to the underestimation of the band gap, they might be placed erroneously
into the conduction band. In such a case, a delocalized conduction band state
and not the localized defect state would be occupied with electrons. In
particular, this is partially true for O~vacancies on TiO$_2$(110). Experiment
indicates that the O~vacancies are responsible for a defect state about
0.7--0.9\,eV below the conduction band edge.\cite{GA84,ExW9,H98}  A careful
survey of the available literature data\cite{GP07}  showed, however, that in
well converged GGA/DFT calculations this defect level is pinned at the bottom
of the conduction band, giving rise to a defect state which is too delocalized.

The problem of both, the band gap and the position of the defect levels,
could be cured to a large extend by employing hybrid density functionals.%
\cite{Ovac3,OvacA,GP07,VP06}  Unfortunately, for extensive studies using
large supercells as required for the present purpose, they are computationally
still very demanding even if localized basis sets are used. On the other hand,
the review Ref.~\onlinecite{GP07}  surprisingly reveals that despite the
problem of the correct position of the defect level, GGA/DFT calculations give
quite good results for the {\em energetics} of the defect formation. While the
best estimate for the O~vacancy formation energy using a cluster model and the
B3LYP hybrid functional is 2.7\,eV (which relies on extrapolations to account
for the finite cluster size and the limited atomic relaxation in this
calculation, thus adding some uncertainty to this value),\cite{GP07}  the 
converged DFT result (using the generalized gradient functionals PW91, PBE
or RPBE) is 3.0\,eV.\cite{GP07}  It is in particular noticeable that despite
the underestimation of the band gap and the too strong delocalization of the
defect state, the DFT calculation {\em overestimates} the vacancy formation
energy and thus underestimate the reducibility of TiO$_2$.

In recent experiments it was shown that the dissociative adsorption of water
at O~vacancies (thereby forming two neighboring OH groups on the surface)
does not change significantly the O~vacancy induced defect state in the band
gap (which was seen 0.9\,eV below the conduction band edge).\cite{HE03}
Also for this hydrogen reduced state of the TiO$_2$(110) surface (which
can be viewed as adsorbing an H$_2$ molecule on the defect-free surface
instead of dissociating a water molecules at an O~vacancy), Di Valenti
{\it et al}.\cite{VP06} found that the defect state is not separated from the
conduction band in GGA/DFT calculations. Nevertheless, despite this problem of
accurately describing the electronic structure, we will argue in the present
paper that the GGA/DFT calculations still give quite reliable results for the
{\em energetics} of reduced surfaces via hydroxylation, similar as we have
seen for the surface reduction via O~vacancy formation. In particular in the
regime of higher H coverages, when the defect states start to interact and a
defect band is formed which will gradually hybridize with the conduction band,
the DFT calculations will be less and less hampered by the band gap problem.

Figure~\ref{F1} shows the atomic structure of the stoichiometric TiO$_2$(110)
surface. In our calculations all surface structures were modeled by
periodically repeated slabs. Slabs for the stoichiometric (110) surface
are built by a stacking sequence of trilayers with a composition of
O--Ti$_2$O$_2$--O. In the bulk, the Ti and O ions are sixfold- and
threefold-coordinated, respectively. At the surface this coordination
is reduced and the characteristic feature of the (110) surface is the
presence of rows of twofold-coordinated bridging O ions parallel to
fivefold-coordinated Ti atom along the [001] direction. Using formal ionic
charges of +4 and $-$2 for the Ti and O ions, respectively, the stoichiometric
(110) surface would be charge neutral. From a more covalent point of view, the
(110) surface, as shown in Fig.~\ref{F1}, is created by a cleavage of the
crystal in which the minimum possible number of bonds has been broken.

The surface energy for the unrelaxed slabs converges very fast with slab
thickness as shown in Fig.~\ref{F2}. After a full relaxation, however,
the surface energy shows strong odd--even oscillations with the number of
trilayers in the slab (see Fig.~\ref{F2}). A very large number of trilayers
would be needed to get well converged results. These oscillations have been
noted previously\cite{DV94,Ovac7,BG04,TL06,HC06,KP07}  and are also present
in plots of vacancy formation\cite{Ovac6,Ovac8,HC06}  and water adsorption
energies.\cite{ThW7,ThW9,ThWA}  The odd--even oscillations arise from the
different symmetry of the slabs which leads to a significant change in the
hybridization of the O 2$p$ and Ti 3$d$ states in the direction of the surface
normal.\cite{BG04}  In slabs with an even number of trilayers the O--Ti
hybridization causes a modulation of the coupling between the trilayers
in such a way that a sequence of pairs of trilayers is formed in which the
coupling of the trilayers within the pair is stronger than the interaction
between pairs.\cite{BG04}  For slabs with an odd number of trilayers, in
contrast, the atomic relaxations, which are associated with this pair
formation, are suppressed by the central Ti$_2$O$_2$ mirror plane and the
O--Ti hybrid orbitals become more delocalized over the whole slab.\cite{BG04}
The surface energy of the slabs with an odd number is higher and the
convergence to the limit of infinite slab thickness is slower than when using
an even number of trilayers. The odd--even oscillations can be largely reduced
if not the full slab is allowed to relax. In Fig.~\ref{F2} the surface energy
is shown for a calculation in which the atoms in the bottom two trilayers were
held fixed at their bulk positions. Using this setup, quite well converged
results are already obtained with only four trilayers.

Based on this observation Thompson and Lewis\cite{TL06} proposed to use such
a slab setup of four trilayers with the bottom two layers fixed for any large
scale calculation. However, the surface properties of a stoichiometric slab
might be less influenced by the slab thickness than when it comes to reduced
surface structures. States above the valence band (either defect states in the
band gap or states from the lower edge of the conduction band) are more
delocalized than the valence states and might be more sensitive to the
truncation at the bottom of the slab. For example, Leconte {\it et al}.%
\cite{LM02} have reported that for fully hydroxylated slabs with a thickness
of two trilayers only 63\% of the spin density is localized at the
hydroxylated top surface layer, but 37\% at the bottom of the slab. The
fully hydroxylated surface is the highest reduced state which we will
consider in our calculations (though, as we will show in Sec.~\ref{Hadsorp},
this configuration is thermodynamically not stable). Even after fixing the
atomic positions of the bottom two trilayers to the bulk positions, the
hydrogen adsorption energy at monolayer coverage still shows a slow
convergence and noticeable oscillations with slab thickness, as can be seen
in Fig.~\ref{F3}. 

Though TiO$_2$ is a strong ionic insulator, the chemical bond has significant
covalent contributions. It is therefore worth to think about the TiO$_2$
surface in terms of broken surface bonds as one would do for covalent
semiconductor surfaces.\cite{LF91}  Ti and O have four and six valence
electrons, respectively. To fulfill the octet rule, each O atom receives
2/3 electrons from its three nearest Ti neighbors and Ti contributes
4/6 electrons to each of its six nearest-neighbor bonds. At the TiO$_2$(110)
surface the fivefold Ti and the twofold bridging O atoms have lost one of
their nearest-neighbor atoms. Thus, dangling bonds are created which are
occupied with 2/3 electrons for the fivefold Ti(5) and 4/3 electrons for the
twofold O(2). At the stoichiometric surface the partial occupation of these
dangling bonds is removed by autocompensation.\cite{LF91}  Alternatively, we
saturate these broken bonds at the bottom of the slab by introducing
artificial atoms with a nucleus charge of +4/3 and +2/3 which we place next
to the Ti(5) and O(2) atoms, respectively, thus creating a more bulk-like
environment for these surface atoms.\cite{MD04}  The distance between the
pseudo-hydrogen and the surface atoms was determined by a geometry
optimization in which all Ti and O atoms were kept fixed at the TiO$_2$ bulk
positions. The effect of saturating the broken surface bonds with the
pseudo-hydrogen atoms on the electronic structure of the surfaces is
illustrated in Fig.~\ref{F4}. Without saturation, the local density of states
(LDOS) of the fixed Ti(5) and O(2) surface atoms (red dashed lines) deviates
strongly from the bulk behavior (black solid lines). The surface band gap is
reduced and strong changes and shifts in the LDOS peaks due to the
autocompensation can be seen. After saturation of the broken surface bonds,
however, the LDOS of the surface atoms (blue dashed-dotted lines) is almost
indistinguishable from the bulk. With this saturation of the bottom of our
slabs we find now also for reduced surface structures a much faster
convergence of surface properties with slab thickness. The oscillations in the
hydrogen adsorption energy in Fig.~\ref{F3} have been reduced from 0.1\,eV to
about 0.02\,eV per adsorbed hydrogen atom. 

Based on these convergence tests we decided to use for all further
calculations slabs with a thickness of four trilayers, including a saturation
of the broken surface bonds at the bottom. The upper two trilayers were always
fully relaxed while the atoms in the lower two layers were fixed at the bulk
positions. A large (4$\times$2) surface unit cell\cite{Com2} was used in order
to be able to study adequately the coverage dependence of hydrogen and water
adsorption. The slab for the stoichiometric surface thus contained 208 atoms,
including the artificial atoms for saturating the broken surface bonds
at the bottom of the slab. The calculated equilibrium bulk values were taken
for the lattice constants parallel to the surface. The slabs were separated by
a vacuum region of about 13\,{\AA} thickness, which corresponds to the
thickness of the slab itself. The surface relaxations of the stoichiometric
surface are essentially the same as described by Thompson and
Lewis\cite{TL06}.  As shown in Fig.~\ref{F2}, the relaxation energy of the
surface is quite substantial. The unrelaxed surface energy is reduced by 
atomic relaxations from 1.64\,eV (1.35\,J/m$^2$) by 1.21\,eV to 0.43\,eV per
surface unit cell (0.36\,J/m$^2$). For comparison, the relaxation energy for
the ZnO(10$\bar{1}$0) surface amounts to only 0.37\,eV per surface unit
cell.\cite{BM03}  As we will see later on, these large surface relaxations
have a strong impact on the coverage dependence of adsorption energies.

In order to analyze the thermodynamic stability of our different surface
structures we assume that the surfaces can exchange O and H atoms with a
surrounding gas phase. Assuming thermodynamic equilibrium, the most stable
surface composition at a given temperature $T$ and pressure $p$ is given by
the minimum of the Gibbs free surface energy $\gamma(T,p)$.\cite{KP87,QM88,%
RS01,BM04}  Since we are only interested in the relative stabilities of
surface structures, we calculate directly the difference $\Delta\gamma(T,p)$
of the Gibbs free surface energy of the defective or adsorbate-covered and the
stoichiometric, ideal surface according to 
\begin{eqnarray}
\label{def_dgamma}
\Delta\gamma(T,p)
 & = & \frac{1}{A} \Big(
       G_{\rm slab}^{\rm surf}(T,p,\Delta N_{\rm O},\Delta N_{\rm H}) -
       G^{\rm ref}_{\rm slab}(T,p) \nonumber \\
 &   & {}+ \Delta N_{\rm O}\,\mu_{\rm O}(T,p) -
       \Delta N_{\rm H}\,\mu_{\rm H}(T,p) \Big) \;,
\end{eqnarray}
where $G_{\rm slab}^{\rm surf}$ and $G_{\rm slab}^{\rm ref}$ are the Gibbs
free energies of the modified and the stoichiometric reference surface
configurations, respectively. $A$ is the surface area, $\Delta N_{\rm O}$,
$\Delta N_{\rm H}$ are the differences in the numbers of O and H atoms between
the two surfaces, and $\mu_{\rm O}(T,p)$, $\mu_{\rm H}(T,p)$ are chemical
potentials representing the Gibbs free energy of the gas phase with which
the O and H atoms are exchanged. According to this definition, $\Delta\gamma$
is negative if the modified surface is thermodynamically more stable than the
stoichiometric surface and positive otherwise. Assuming that all differences
in entropy and volume contributions in $\Delta\gamma$ are
negligible,\cite{RS01,BM04} we approximate the Gibbs free energies
$G_{\rm slab}^{\rm surf}$ and $G_{\rm slab}^{\rm ref}$ by their respective
total energies of our DFT slab calculations as usual.\cite{RS01,BM04}  Upper
bounds for the chemical potentials $\mu_{\rm O}$ and $\mu_{\rm H}$ are given
by the total energies of their most stable elemental phases,\cite{QM88}
that is, molecular oxygen $\frac{1}{2}E_{\rm mol}^{{\rm O}_2}$ and molecular
hydrogen $\frac{1}{2}E_{\rm mol}^{{\rm H}_2}$, respectively. These upper
bounds are taken as new zero point of energy by introducing $\Delta\mu_{\rm O}
= \mu_{\rm O} - \frac{1}{2}E_{\rm mol}^{{\rm O}_2}$ and $\Delta\mu_{\rm H} = 
\mu_{\rm H} - \frac{1}{2}E_{\rm mol}^{{\rm H}_2}$. A lower bound for
$\Delta\mu_{\rm O}$ is given by minus half of the formation energy of bulk
TiO$_2$, i.e. $E_{\rm f}^{{\rm TiO}_2} = E_{\rm bulk}^{\rm Ti} +
E_{\rm mol}^{{\rm O}_2} - E_{\rm bulk}^{{\rm TiO}_2}$
(here $E_{\rm bulk}^{{\rm TiO}_2}$ and $E_{\rm bulk}^{\rm Ti}$ are the
energies of one bulk unit cell of TiO$_2$ and metallic Ti, respectively),%
\cite{RS01,BM04}  for which we have taken the theoretical value of 4.80\,eV
from our PBE/DFT calculations. The chemical potential can be related to
experimental temperature and pressure conditions by using experimental
thermochemical reference data or by applying the ideal gas equation.%
\cite{RS01,BM04,BA00}

\begin{table}
\caption{\label{T1}
Vacancy formation energies $E_{\rm v}$ (in eV) for different oxygen vacancy
configurations on TiO$_2$(110) in the limit of $\Delta\mu_{\rm O}=0$
(oxidizing, i.e. oxygen rich conditions).}
\begin{ruledtabular}
\begin{tabular}{ccc}
Vacancy type & $E_{\rm v}$ singlet & $E_{\rm v}$ triplet\\
\hline
O(2)--v    &  3.07  &  3.02 \\
O(3)--v    &  4.05  &  4.00 \\
TiO$_2$--v &  1.03
\end{tabular}
\end{ruledtabular}
\end{table}

The transition state search for the dissociation, desorption and migration
processes was conducted with the nudged elastic band (NEB)\cite{JM98} and
the dimer method.\cite{HJ99,HJ0a,HJ0b,HB05}  Fully relaxed configurations
were chosen as initial and final state of the NEB calculations. Throughout
twelve images were used which were connected by springs with a fixed spring
constant of about 20\,eV/{\AA}$^2$. The two images in the dimer calculations
were separated by 0.01\,{\AA} in configuration space and the trial steps
for translation and rotation were 0.01\,{\AA} and 10$^\circ$, respectively.
In all calculations first a good approximation of the transition path was
determined with NEB which was then refined with a dimer method run for a
precise location of the transition state.

%------------------------------------------------------------------------------
\section{Results and Discussion}
%------------------------------------------------------------------------------

\subsection{Oxygen Vacancies on TiO$_2$(110)}

We have investigated three different types of oxygen-related defects.
Vacancies are created by removing a single bridging oxygen O(2) atom,
a threefold-coordinated O(3) atom from the surface layer, or a full TiO$_2$
unit of two neighboring bridging O(2) atoms and the underlying Ti cation,
see Fig.~\ref{F1}. With our choice of (4$\times$2) supercells this corresponds
to a vacancy concentration of 1/8 monolayer and a defect separation of about
12\,{\AA}. The vacancy formation energies for the missing O(2) and O(3) atoms
and the TiO$_2$ unit are given by\cite{KM07}
\begin{equation}
E_{\rm v}^{\rm O} = E_{\rm slab}^{\rm O-v} +
\frac{1}{2} E_{\rm mol}^{{\rm O}_2} - E_{\rm slab}^{\rm ref} +
\Delta\mu_{\rm O}
\end{equation}
and
\begin{equation}
E_{\rm v}^{{\rm TiO}_2} = E_{\rm slab}^{{\rm TiO}_2-{\rm v}} +
E_{\rm bulk}^{{\rm TiO}_2} - E_{\rm slab}^{\rm ref} \;,
\end{equation}
respectively, where $E_{\rm slab}^{\rm O-v}$,
$E_{\rm slab}^{{\rm TiO}_2-{\rm v}}$ and $E_{\rm slab}^{\rm ref}$ are the
total energies for the defective and the stoichiometric slab, respectively.
The results in Table~\ref{T1} show that the O(2) and O(3) vacancies are
slightly more stable in the triplet than in the singlet state, in agreement
with previous calculations.\cite{Ovac2,Ovac3,Ovac6,Ovac8}  As one would expect
from the local coordination, the formation of an O(2) defect is much more
favorable than the O(3) vacancy. Our value for the O(2) vacancy formation
energy of 3.02\,eV is basically identical to the results of Rasmussen
{\it et al}.\cite{Ovac6} (3.03\,eV) and Oviedo {\it et al}.\cite{Ovac8}
(3.07\,eV, singlet state) who have both conducted extensive studies on the
convergence of $E_{\rm v}^{O(2)}$ with defect concentration, slab thickness
and the number of relaxed surface layers. Also Ganduglia-Pirovano
{\it et al}.\cite{GP07} came to the conclusion in their survey of the
available literature data that the converged value for the formation energy
of isolated O(2) vacancies in GGA/DFT calculations (using the PW91, PBE or RPBE
functional) is about 3.0\,eV. This supports that our choice of slab thickness,
relaxation, bond saturation at the bottom of the slab, and size of surface
unit cell is indeed appropriate to give well converged results.

\begin{figure}[!t]
\noindent
\includegraphics[width=246pt]{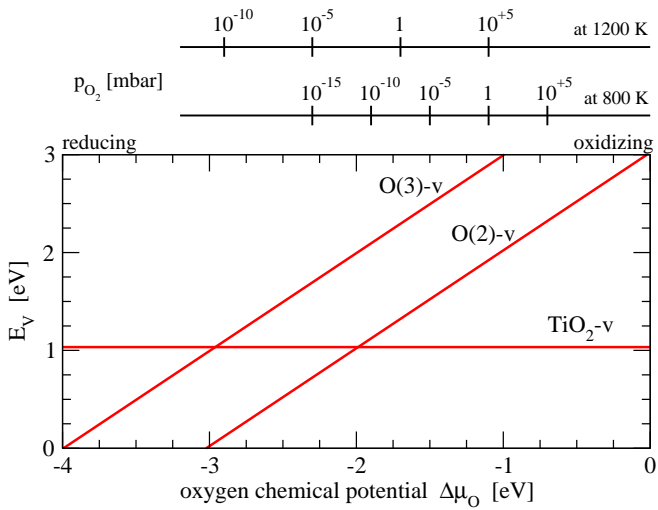}
\caption{\label{F5}
Defect formation energy $E_{\rm v}$ for O~vacancies at the TiO$_2$(110)
surface as function of the oxygen chemical potential $\Delta\mu_{\rm O}$.
In the top $x$-axis, the chemical potential has been translated into a
partial pressure scale $p_{\rm O_2}$ of molecular oxygen at 800\,K and
1200\,K.}
\end{figure}

In Fig.~\ref{F5} the vacancy formation energies are plotted as function of
the oxygen chemical potential $\Delta\mu_{\rm O}$ of the environment. Assuming
thermodynamic equilibrium of the surfaces with an O$_2$ gas phase we have
converted the chemical potential into a pressure scale for a temperature
of 800\,K and 1200\,K using the ideal gas equation.\cite{RS01,BM04}  Since we
have neglected the entropy contributions to the Gibbs free energy of the
defective and stoichiometric slabs, we cannot capture with our description
the formation of {\it thermal} oxygen vacancies which are due to the gain in
configurational entropy by the random distribution of the defects on the
surface. Therefore, at lower temperatures we will always find the
stoichiometric, defect-free surface to be the most stable one. We can only
predict the appearance of {\it structural} vacancies which are formed if
the gain in entropy by bringing oxygen into the gas phase outweighs
the binding energy to the surface. Having this in mind we have to expect
that we will overestimate the temperature at which significant amounts
of O~vacancies (in the percent range) will form. From Fig.~\ref{F5} we
see that at UHV conditions (base pressure of about 10$^{-10}$\,mbar) the
onset for the formation of structural O(2) vacancies is around 1200\,K
(i.e., the chemical potential at which the vacancy formation energy
$E_{\rm v}$ becomes zero). This is in quite good agreement with 
experimental observations,\cite{DB03} taking into account that an
uncertainty of the vacancy formation energy of a few tenths of an eV
from the DFT calculations translates into an error bar for this temperature
in the range of 100--200\,K. Rasmussen {\it et al}.\cite{Ovac6}  and
Oviedo {\it et al}.\cite{Ovac8}  have shown that the interaction between
the O(2) vacancies is strongly repulsive and that the O~vacancy formation
energy increases rapidly with higher defect concentrations. Thus, structures
with higher O defect concentrations will appear farther to the left in the
phase diagram Fig.~\ref{F5} so that increasingly reducing conditions are
needed to reach defect concentrations beyond a few percent.

On the other hand, quite surprisingly, we find that the formation energy for
the TiO$_2$ defects is much lower than for the O(2) and O(3) vacancies over a
wide range of the O chemical potential. Already at temperatures around 800\,K
(at UHV conditions) it becomes thermodynamically more favorable to convert the
O(2) vacancies into TiO$_2$ defects. A similar result was recently obtained
for the ZnO(10$\bar{1}$0) surface,\cite{KM07}  where it was found that ZnO
dimer vacancies are thermodynamically more stable than O~vacancies for most
experimental conditions. While for ZnO a strong suppression of O~vacancies
on the (10$\bar{1}$0) surface can be expected, since Zn has a very low vapor
pressure and can easily desorb from the surface (indeed, desorption of Zn is
regularly observed in TDS experiments of ZnO), it is more difficult in the
case of TiO$_2$ to remove the Ti ions from the surface since the kinetics of
the process will play a much larger role. However, one has to consider that
the O~vacancies, created on the TiO$_2$(110) surface in UHV experiments by
annealing, sputtering or electron bombardment, are only present due to kinetic
limitations, and they will be eliminated under ambient conditions not only by
a dissociative adsorption of O$_2$ and water, but there is also a strong
driving force to transform them into TiO$_2$ vacancies, for example, by a
mechanism in which Ti interstitials are created.\cite{WB08}

\begin{table}
\caption{\label{T2}
Adsorption energies $E_{\rm ad}^{\rm H}$ of H atoms (in eV/atom) on
TiO$_2$(110). Configurations with negative values of $E_{\rm ad}^{\rm H}$
are energetically unstable towards the desorption of H$_2$ molecules.
$N_{\rm H}$ is the number of hydrogen atoms in the (4$\times$2) surface unit
cell. The arrangements of the H atoms on the surface are illustrated in
Fig.~\ref{F6}.}
\begin{ruledtabular}
\begin{tabular}{clr@{\hspace*{8pt}}|clr}
$N_{\rm H}$  &  Config  &  E$_{\rm ad}^{\rm H}$  &
$N_{\rm H}$  &  Config  &  E$_{\rm ad}^{\rm H}$ \\
\hline
 1  & O(2)       &    0.563  &   5  & 5H          &    0.229 \\
    & O(3)       & $-$0.062  &      & \\
    & Ti(5)      & $-$2.217  &   6  & 6H$-$a      &    0.143 \\
 2  & O(2)+O(2)  &    0.400  &      & 6H$-$b      &    0.127 \\
    & O(2)+O(3)  &    0.142  &      & \\
    & O(2)+Ti(5) & $-$0.147  &   8  & 8O(2)       &    0.033 \\
 4  & 4H$-$a     &    0.280  &      & 7O(2)+O(3)  & $-$0.007 \\
    & 4H$-$b     &    0.215  &      & 7O(2)+Ti(5) & $-$0.013 \\
    & 4H$-$c     &    0.211  &      & 7O(2)+O$_{\rm sub}^{\rm a}$ & $-$0.009 \\
    & 4H$-$d     &    0.202  &      & 7O(2)+O$_{\rm sub}^{\rm b}$ & $-$0.012 \\
    & 4H$-$e     &    0.128  &      & \\
\end{tabular}
\end{ruledtabular}
\end{table}

\begin{figure}[!t]
\noindent
\includegraphics[width=246pt]{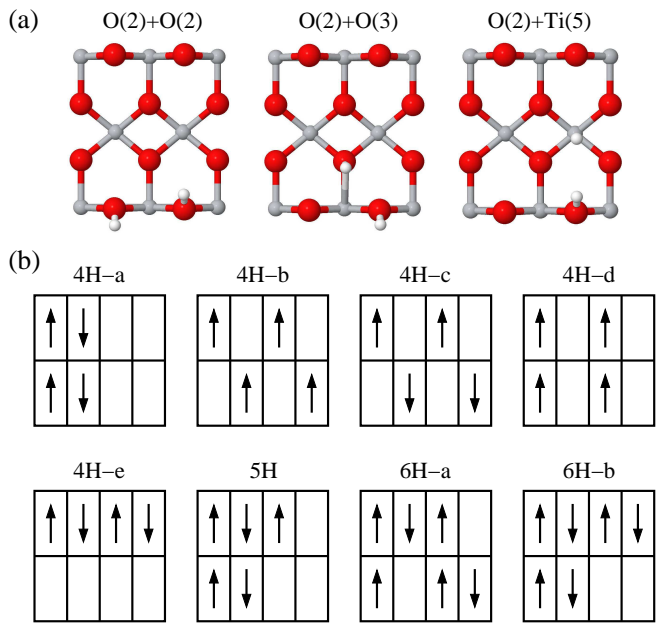}
\caption{\label{F6}
Schematic illustration of some configurations of the TiO$_2$(110) surface
with adsorbed H atoms. (a) Configurations with two adsorbed H atoms. Ti,
O and H atoms are depicted as small gray, large red and small white spheres,
respectively. (b) Configurations with higher H coverage. The arrows indicate
the tilt of the OH groups after H adsorption. The corresponding adsorption
energies $E_{\rm ad}$ are given in Table~\ref{T2}.}
\end{figure}

%------------------------------------------------------------------------------

\subsection{Hydrogen Adsorption on TiO$_2$(110)}
\label{Hadsorp}

In order to investigate the interaction of hydrogen with the TiO$_2$(110)
surface we have calculated the total energies of various hydroxylated
TiO$_2$(110) surface structures taking into account both, different hydrogen
coverages and adsorption sites for the H atoms. Our results for the hydrogen
adsorption energies $E_{\rm ad}^{\rm H}$ per H atom
\begin{equation}
E_{\rm ad}^{\rm H} = \frac{1}{N_{\rm H}} \Big(
E_{\rm slab}^{\rm ref} + \frac{N_{\rm H}}{2} E_{\rm mol}^{{\rm H}_2} -
E_{\rm slab}^{\rm H-ad}(N_{\rm H}) \Big) \;,
\end{equation}
where $E_{\rm slab}^{\rm H-ad}(N_{\rm H})$ is the total energy of the
slab calculation with $N_{\rm H}$ adsorbed H atoms, are summarized in
Table~\ref{T2}. Single H atoms adsorb preferentially ontop of the bridging
O(2) atoms, whereas the O(3) and Ti(5) sites are significantly higher in
energy. The OH groups, which form upon H adsorption, do not remain upright
but break the mirror symmetry of the surface by tilting about 20$^\circ$
(isolated OH groups) and up to 50$^\circ$ (hydroxylation of all bridging O
atoms) with respect to the surface normal.

For pairs of H atoms we find that the homolytic adsorption on two
neighboring O(2) sites (thereby reducing the surface) is more stable than
the heterolytic adsorption on O(2) and Ti(5) (for the atomic configurations,
see Fig.~\ref{F6}a). The heterolytic adsorption is even energetically unstable
towards desorption of H$_2$. This preference of the homolytic adsorption
of hydrogen is expected for an easily reducible oxide. Table~\ref{T2} also
shows that the interaction between neighboring OH groups is repulsive. The
adsorption energy per H atom decreases from 0.56\,eV for isolated OH groups
to 0.40\,eV for neighboring OH pairs.

By comparing the energy of the surface structure with two neighboring OH
groups to the structure with an O(2) vacancy, we can directly deduce the
energy gain for dissociatively adsorbing a water molecule at isolated O(2)
defects by applying a thermodynamic cycle. For the water adsorption energy
at O(2) vacancies we find a value of 1.31\,eV, which is slightly larger
than previously reported results (0.94, 0.97 and 1.10\,eV according to
Refs.~\onlinecite{ExW2,TV05,ExW3}, respectively), but agrees well with an
estimated value of about 1.4\,eV deduced by a Redhead analysis\cite{RH62}
of a water desorption peak at 520\,K in TDS experiment which was assigned
to the recombination of neighboring OH groups.\cite{HE03}  According to
our NEB calculations this process is without barrier\cite{Com3} so that
the activation energy for desorption is given by the adsorption energy.
Since the previous calculations of Refs.~\onlinecite{ExW2,TV05,ExW3} were
done with (2$\times$2) surface unit cells for which there is still significant
interaction between O~vacancies and pairs of hydroxyl groups (which can
be seen, for example, from our results in Table~\ref{T2}), we attribute
our slightly higher value for the binding energy of water at O~vacancies
to our larger (4$\times$2) surface unit cell which represent better converged
results with respect of describing isolated defects and OH pairs. It is
interesting to see that we underestimate the binding energy of water in
the O~vacancies. Since the PBE/DFT calculations tend to overestimate the 
O~vacancy formation energy,\cite{GP07}, this means that also the energy of
the reduced surface via hydroxylation is slightly too high. Thus, as in the
case of surface reduction via O depletion, also the reducibility of TiO$_2$
via hydrogen adsorption is {\it underestimated} in the PBE/DFT calculations.

\begin{figure}[!t]
\noindent
\includegraphics[width=246pt]{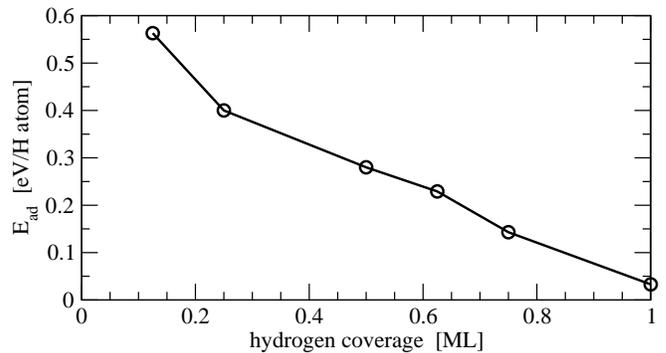}
\caption{\label{F7}
Hydrogen adsorption energy $E_{\rm ad}$ as function of the hydrogen coverage
of the surface.}
\end{figure}

With increasing hydrogen coverage, O(2) remains the preferred adsorption site
up to the coverage of a full monolayer. No significant amounts of H atoms
will be adsorbed on O(3) and Ti(5) sites. Neighboring OH groups in the [001]
direction tend to tilt in opposite directions (see Fig.~\ref{F6}a), whereas
the coupling between [001] rows (concerning the tilt direction of the OH
groups) is relatively weak (with a minor preference of the same tilt direction
of OH groups in [1$\bar{1}$0] rows). The ground state configuration of the
fully hydroxylated surface, therefore, will have a (2$\times$1) symmetry.
However, since the barrier for flipping the orientation of an OH group is
very small, they will be fully orientally disordered at room temperature.

\begin{figure}[!t]
\noindent
\includegraphics[width=246pt]{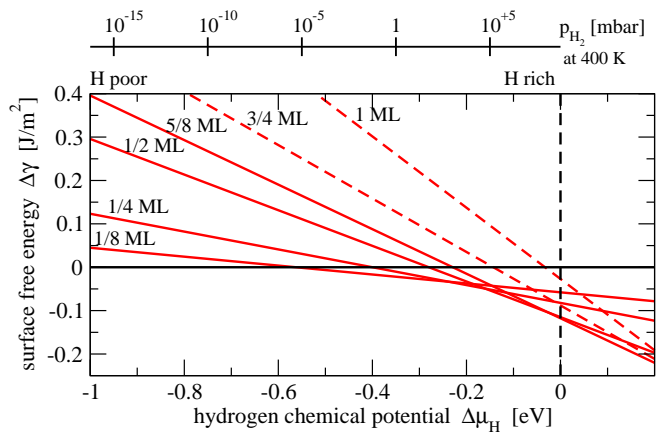}
\caption{\label{F8}
Surface Gibbs free energy $\Delta\gamma$ of TiO$_2$(110) surfaces with
different hydrogen coverages as function of the hydrogen chemical potential
$\Delta\mu_{\rm H}$. In the top $x$ axis, the chemical potential has been
translated into a partial pressure scale $p_{\rm H_2}$ of molecular hydrogen
at 400\,K. Thermodynamically unstable surface structures are represented by
dashed lines.}
\end{figure}

Though O(2) remains to be the preferred adsorption site also with increasing
H coverage, the adsorption energy $E_{\rm ad}^{\rm H}$ per H atom, i.e. the
feasibility to further reduce the TiO$_2$(110) surface, decreases strongly
upon adsorption of more hydrogen (see Fig.~\ref{F7}). At full monolayer
coverage $E_{\rm ad}^{\rm H}$ is still positive. However, if the interaction
energy of hydrogen with the TiO$_2$(110) surface is not expressed as binding
energy per H atom but as energy gain per surface area it becomes obvious from
the phase diagram Fig.~\ref{F8} that the fully hydroxylated surface is
thermodynamically unstable. The Gibbs free surface energy $\Delta\gamma$
for a hydrogen monolayer (this also applies for the 0.75\,ML coverage)
is higher than the surface energies of lower coverages over the whole range
of the hydrogen chemical potential. I.e., the energy of the fully hydroxylated
surface can be always lowered by desorbing H$_2$ molecules and reducing the
H coverage. The highest coverage that can be reached in thermodynamic
equilibrium in the limit of zero hydrogen chemical potential (i.e. at low
temperature and high hydrogen partial pressure) is around 60\%--70\%. This
is in excellent agreement with recent experimental observations where by
exposing the TiO$_2$(110) surface to atomic hydrogen no H coverage beyond 70\%
could be obtained.\cite{YC08}  The experimental setup of exposing the surface
to atomic hydrogen is an almost ideal realization of the theoretical
assumption underlying the calculation of the phase diagram Fig.~\ref{F8},
namely that the surface is in thermodynamic equilibrium with a surrounding
H$_2$ gas phase: no kinetic limitations are present and H atoms can adsorb
and desorb without encountering any barrier.

In conclusion, this remarkable agreement between theory and experiment
suggests that despite the problem of the band gap and the correct position of
the defect levels (which we analyzed and discussed in Sec.~\ref{method}), the
PBE/DFT calculations seem to be quite able to capture accurately the subtle
balance between the energy cost to dissociate H$_2$ molecules versus that of
reducing the TiO$_2$(110) surface.

%------------------------------------------------------------------------------

\subsection{Migration of Hydrogen Atoms}

Upon heating the hydroxylated TiO$_2$(110) surface a quite unexpected
behavior was observed.\cite{KB04,YC08}  Monitoring the He atom reflectivity
of the surface revealed two distinct maxima at 388\,K and 625\,K (with a small
shoulder at 560\,K) at which the surface undergoes structural rearrangements.
These rearrangements must involve the loss of H atoms since the surface is
H--free above 650\,K. When the Redhead formula\cite{RH62,PG01} with a frequency
prefactor of 10$^{-13}$\,s$^{-1}$ is applied, the two temperatures can be
translated (with an uncertainty of about 20\%) into activation barriers of
1.03\,eV and 1.68\,eV, respectively, for the processes which are involved in
the structural changes. In a first assignment the two processes were assumed
to be the desorption of hydrogen from the Ti(5) and O(2) sites\cite{KB04},
respectively. Based on our results on the hydrogen adsorption energy for the
different surface sites, however, this interpretation can be clearly ruled
out. Surprisingly, in a subsequent conventional TDS measurement, no desorption
of H$_2$ at all and only a very small amount of water could be detected. This
led the authors to the conclusion that instead of desorption taking place,
the H atoms migrate into the bulk.\cite{YC08}

\begin{table}
\caption{\label{T3}
Activation energy $E_{\rm act}$ (in eV) for different migration paths
of H atoms on the hydroxylated TiO$_2$(110) surface. All calculations
are based on a fully hydroxylated surface with eight H atoms in the
(4$\times$2) surface unit cell. The temperature $T_{\rm act}$ (in K),
at which the onset of the processes can be expected, has been estimated
from the activation energy $E_{\rm act}$ by applying a standard Redhead
analysis\cite{RH62,PG01} with a frequency factor of 1$\cdot$10$^{13}$~s$^{-1}$
and a heating rate of 1~K/s.}
\begin{ruledtabular}
\begin{tabular}{llc}
Migration Path  &  $E_{\rm act}$  &  $T_{\rm act}$ \\
\hline
O(2) $\longrightarrow$ O$_{\rm sub}^{\rm a}$  &  2.56  &  $\rm 963 \pm 197$\\
O(3) $\longrightarrow$ O$_{\rm sub}^{\rm a}$  &  0.93  &  $\rm 361 \pm  72$\\
Ti(5) $\longrightarrow$ O$_{\rm sub}^{\rm a}$ &  1.52  &  $\rm 581 \pm 117$\\
O(2) $\longrightarrow$ O(3)                   &  0.63  &  $\rm 248 \pm  49$\\
O(2) $\longrightarrow$ Ti(5)                  &  1.54  &  $\rm 588 \pm 119$\\
O(2) $\longrightarrow$ O(2)                   &  1.21  &  $\rm 466 \pm  93$\\
2*O(2) $\longrightarrow$ H$_2$(gas)           &  1.85  &  $\rm 703 \pm 143$\\
\end{tabular}
\end{ruledtabular}
\end{table}

Aiming at identifying the processes which lead to the structural changes of
the surface and the loss of the H atoms we have calculated the activation
barriers for different migration paths of H atoms on the hydroxylated
TiO$_2$(110) surface. Some of these migration paths have been considered
already in a previous DFT study by Yin {\it et al.},\cite{YC08}  in which,
however, only the smallest possible (1$\times$1) surface unit cell with two
adsorbed H atoms was used. In our calculations we start from the fully
hydroxylated surface with eight H atoms adsorbed ontop of the bridging O(2)
atoms in a (4$\times$2) surface unit cell. In the first step we examine how
much the energy increases if one of the H atoms from the O(2) sites is placed
ontop O(3), Ti(5), or subsurface. As we can see from Table~\ref{T2} all three
sites are energetically almost degenerate and only about 0.3\,eV higher in
energy than the O(2) site (note that the values in Table~\ref{T2} are
adsorption energies per H atom. Therefore, to get the energy difference
between two structures with different arrangement of the H atoms, the
$E_{\rm ad}^{\rm H}$ values have to be multiplied by the number of adsorbed
H atoms, i.e. eight, before subtracting the energies). This energy difference
between the adsorption sites increases when the H coverage decreases. At
1/4 monolayer H coverage the O(3) and Ti(5) sites are higher in energy by
0.52\,eV and 1.09\,eV than the O(2) site, respectively (see Table~\ref{T2}).

Returning back to the fully hydroxylated surface we find that the transition
barrier for migration of an H atom directly from an O(2) site to a subsurface
O atom is quite high (see Table~\ref{T3}), whereas diffusion to the O(3) and
Ti(5) sites is much more favorable. On the other hand, the activation energies
for transferring an H atom from O(3) and Ti(5) to a subsurface O atom are
moderate. Overall, the energetically most favorable migration path is from
O(2) via O(3) to a subsurface site with barriers of only 0.63\,eV and
0.93\,eV, respectively. In contrast, we find for the desorption of H atoms
from two neighboring O(2) sites via recombination and formation of H$_2$
molecules an activation barrier of 1.85\,eV. This implies that it is indeed
much easier for the H atoms to diffuse into the bulk instead of desorbing
from the surface. The activation barrier of 0.93\,eV also corresponds nicely
to the temperature of 388\,K at which the onset of structural changes of the
hydroxylated TiO$_2$(110) surface was observed in the HAS experiments.%
\cite{KB04}  Altogether our results fully support the conclusions of
Yin {\it et al.}\cite{YC08}  that H atoms from a hydroxylated TiO$_2$(110)
surface do not desorb but prefer to migrate into the bulk instead.

\begin{table}
\caption{\label{T4}
Literature survey of calculated water adsorption energies $E_{\rm ad}^{\rm w}$
(in eV/molecule) at monolayer coverage for molecular (`mol'), mixed
dissociative/molecular (`mix') and dissociative (`diss') adsorption on
TiO$_2$(110).}
\begin{ruledtabular}
\begin{tabular}{llccc}
Functional  &  Reference  &  $E_{\rm ad}^{\rm w}$ mol  &
$E_{\rm ad}^{\rm w}$ mix  &  $E_{\rm ad}^{\rm w}$ diss \\
\hline
BP86\cite{B88,P86} & Ref.~\onlinecite{ThW1,ThW2} &  0.82  &        &  1.08 \\
PW91\cite{PW91}    & Ref.~\onlinecite{ThW3}      &  1.13  &        &  0.98 \\
PW91\cite{PW91}    & Ref.~\onlinecite{ThW4}      &  0.99  &  1.10  &  0.91 \\
RPBE\cite{RPBE}    & Ref.~\onlinecite{ThW7}      &        &  0.64  &  0.37 \\
RPBE\cite{RPBE}    & Ref.~\onlinecite{ThW8}      &  0.52  &  0.53  &  0.34 \\
PW91\cite{PW91}    & Ref.~\onlinecite{ThW9}      &  1.09  &  1.05  &  0.91 \\
PW91\cite{PW91}    & Ref.~\onlinecite{ThWA}      &  1.01  &  0.95  &  0.90 \\
PBE\cite{PBE}      & this work                   &  0.82  &  0.77  &  0.63 \\
\end{tabular}
\end{ruledtabular}
\end{table}

\begin{table}
\caption{\label{T5}
Literature survey of calculated water adsorption energies $E_{\rm ad}^{\rm w}$
(in eV/molecule) of isolated molecules for molecular (`mol') and dissociative
(`diss') adsorption on TiO$_2$(110). The size of the surface unit cell used in
the respective calculation is given in the second column. The (2$\times$2) and
c(4$\times$2) cells correspond to an effective water coverage of 1/4
monolayer, whereas a (4$\times$2) cell represents a 1/8 monolayer water
coverage.}
\begin{ruledtabular}
\begin{tabular}{lclccc}
Functional  & unit cell  &  Reference  &  $E_{\rm ad}^{\rm w}$ mol  &
$E_{\rm ad}^{\rm w}$ diss & $\Delta E_{\rm ad}$ \\
\hline
RPBE\cite{RPBE} & (2$\times$2) & Ref.~\onlinecite{ExW2} &  0.56  &$-$0.23 &  +0.79 \\
RPBE\cite{RPBE} & (2$\times$2) & Ref.~\onlinecite{ThW8} &  0.36  &  0.56  &$-$0.20 \\
PW91\cite{PW91} & (2$\times$2) & Ref.~\onlinecite{ThW9} &  0.83  &  0.71  &  +0.12 \\
PBE\cite{PBE}   &c(4$\times$2) & Ref.~\onlinecite{TV05} &  0.76  &  0.66  &  +0.10 \\
RPBE\cite{RPBE} & (2$\times$2) & Ref.~\onlinecite{ExW3,ExW4} &  0.66  &  0.79  &$-$0.13 \\
PBE\cite{PBE}   & (2$\times$2) & this study             &  0.82  &  0.75  &  +0.07 \\
PBE\cite{PBE}   & (4$\times$2) & this study             &  0.93  &  1.04  &$-$0.11 \\
\end{tabular}
\end{ruledtabular}
\end{table}

%------------------------------------------------------------------------------

\subsection{Water Adsorption on TiO$_2$(110)}

From the experimental point of view there is a wide consensus that water
adsorbs molecularly on defect-free terraces of the TiO$_2$(110) surface.%
\cite{DB03,ExW6,ExW7,ExW8}  In XPS\cite{ExW6} and HREELS\cite{ExW7}
measurements only small amounts of dissociated water could be detected
under UHV conditions, which may be naturally linked to the presence of
residual defects. In TDS\cite{ExW6,ExW7,ExW8}  an intense desorption peak
is observed at 270\,K which could be attributed to the desorption of a
full monolayer of molecularly adsorbed water molecules at Ti$^{4+}$ sites
based on XPS, HREELS and work function measurements. Interestingly, the
peak position shifts to higher temperatures if the initial water coverage
$\Theta$ is lowered, indicating a repulsive interaction between the adsorbate
molecules. This is unusual since for water molecules one would expect an
attractive interaction due to the formation of hydrogen bonds, but the same
behavior is also observed for the anatase (101) surface.\cite{HD03}  The TDS
data could be best described by fitting them to a first-order kinetics model.
Activation energies for desorption of (0.74$-$0.09\,$\Theta$)\,eV\cite{ExW6}
and (0.73$-$0.07\,$\Theta$)\,eV\cite{ExW7} were obtained, whereas from a
modulated molecular beam study\cite{ExW8} a value of
(0.83$-$0.36\,$\Theta$)\,eV was deduced.

\begin{figure}[!t]
\noindent
\includegraphics[width=246pt]{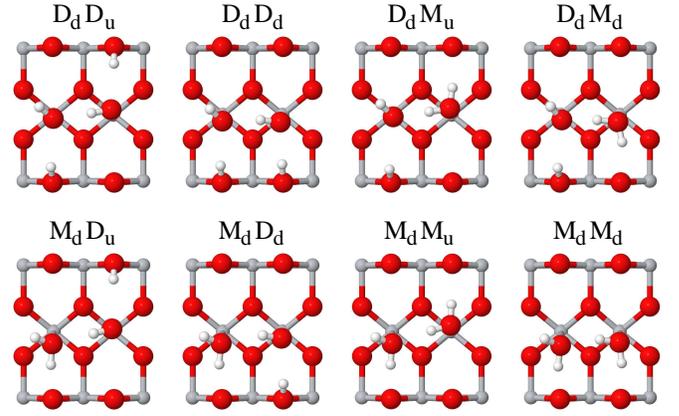}
\caption{\label{F9}
Schematic illustration of water pairs adsorbed on the TiO$_2$(110) surface.
Ti, O and H atoms are depicted as small gray, large red and small white
spheres, respectively. The labels are defined in Table~\ref{T6}.}
\end{figure}

On the other hand, over the past years theoretical studies on the adsorption
behavior of water on TiO$_2$(110) came to very contradictory conclusions.
In early GGA/DFT studies a much lower adsorption energy for molecular than
for dissociative adsorption\cite{ThW1,ThW2} (at full monolayer coverage) was
obtained (it should be noted, however, that the binding energy for the
molecular adsorption was underestimated since only a symmetric adsorption of
the molecules was considered, which is not the most stable geometry, see
Fig.~\ref{F9}). Subsequently, Bates {\it et al}.\cite{ThW3} found a slightly
higher adsorption energy for molecular monolayers, whereas Lindan
{\it et al.}\cite{ThW4,ThW7,ThW8}  proposed a mixed molecular/dissociated
structure as the most stable state. In a Car--Parrinello molecular dynamics
study\cite{ThW6}  it was observed that water does not dissociate on the
perfect surface, suggesting that the dissociation is hindered by a larger
dissociation barrier. Later static calculations of the activation energy for
water dissociation supported this point of view.\cite{ThW8}  In a quite recent
study, however, again molecular adsorption was found to be the most stable
adsorption mode not only for water monolayers but even at lower water
coverages.\cite{ThW9}

\begin{table}
\caption{\label{T6}
Water adsorption energies $E_{\rm ad}^{\rm w}$ (in eV/molecule) on
TiO$_2$(110) for different coverages taking into account molecular(M)
and dissociated (D) states of the water molecules with downward ($d$)
or upward ($u$) orientation. $N_{\rm w}$ is the number of water molecules
in the (4$\times$2) surface unit cell. Parenthesis denote pairs of water
molecules as illustrated in Fig.~\ref{F9}. The structures with four water
molecules consist of two water pairs filling one [001] row and leaving every
second one empty. At full monolayer coverage, both [001] rows are equally
filled with water pairs. The periodicity of the resulting structure is
reported in the `Config' column.}
\begin{ruledtabular}
\begin{tabular}{clr@{\hspace*{8pt}}|clr}
$N_{\rm w}$ & Config & E$_{\rm ad}^{\rm w}$ &
$N_{\rm w}$ & Config & E$_{\rm ad}^{\rm w}$ \\
\hline
 1 & \multicolumn{2}{l|}{single water}
                            &  4 & \multicolumn{2}{l}{double pairs along [001]} \\
   & D$_d$          &  1.04 &    & 2\,(D$_d$\,D$_u$)--(2$\times$2) &  0.67 \\
   & M$_d$          &  0.93 &    & 2\,(D$_d$\,D$_d$)--(1$\times$2) &  0.64 \\
 2 & \multicolumn{2}{l|}{single water}
                            &    & 2\,(D$_d$\,M$_u$)--(2$\times$2) &  0.81 \\
   & 2\,D$_d$ -- c(4$\times$2)
                    &  0.97 &    & 2\,(D$_d$\,M$_d$)--(2$\times$2) &  0.80 \\
   & 2\,D$_d$ -- p(4$\times$1)
                    &  0.96 &    & 2\,(M$_d$\,M$_u$)--(2$\times$2) &  0.84 \\
   & 2\,D$_d$ -- p(2$\times$2)
                    &  0.75 &    & 2\,(M$_d$\,M$_d$)--(1$\times$2) &  0.86 \\
   & 2\,M$_d$ -- c(4$\times$2)
                    &  0.83 &    & \\
   & 2\,M$_d$ -- p(4$\times$1)
                    &  0.82 &    & \\
   & 2\,M$_d$ -- p(2$\times$2)
                    &  0.80 &    & \\
   & water pairs    &       &  8 & \multicolumn{2}{l}{quadruple pairs} \\
   & (D$_d$\,D$_u$) &  0.94 &    & 4\,(D$_d$\,D$_u$)--(2$\times$1) &  0.61 \\
   & (D$_d$\,D$_d$) &  0.90 &    & 4\,(D$_d$\,D$_d$)--(1$\times$1) &  0.63 \\
   & (D$_d$\,M$_u$) &  0.97 &    & 4\,(D$_d$\,M$_u$)--(2$\times$1) &  0.77 \\
   & (D$_d$\,M$_d$) &  0.96 &    & 4\,(D$_d$\,M$_d$)--(2$\times$1) &  0.77 \\
   & (M$_d$\,D$_u$) &  0.90 &    & 4\,(M$_d$\,M$_u$)--(2$\times$1) &  0.80 \\
   & (M$_d$\,D$_d$) &  0.88 &    & 4\,(M$_d$\,M$_d$)--(1$\times$1) &  0.82 \\
   & (M$_d$\,M$_u$) &  0.88 &    & \\
   & (M$_d$\,M$_d$) &  0.90 &    & \\
\end{tabular}
\end{ruledtabular}
\end{table}

The binding energies $E_{\rm ad}^{\rm w}$ from these studies for molecular,
mixed molecular/dissociative, and full dissociative adsorption at monolayer
coverage are summarized in Table~\ref{T4}. The situation becomes even more
confused if the published results for the adsorption of `isolated' water
molecules are compared (see Table~\ref{T5}). A survey of the publications
shows that the results for the water binding energies $E_{\rm ad}^{\rm w}$ 
depend sensitively on the computational setup, in particular the slab
thickness and how the results are extrapolated to the limit of infinite thick
slabs, as well as on the GGA functional which was used in the calculations
(though the result reported in Ref.~\cite{ExW2} on the dissociative water
adsorption is probably due to an error in the calculations since it was
never reproduced in more recent studies). From embedded cluster calculations
using wave function based methods (Hartree--Fock, B3LYP and MP2) it has been
suggested that overall GGA/DFT might overestimate the stability of the
dissociated state of the water molecules compared to a molecular
adsorption.\cite{ThW5}  However, embedded cluster calculations have to be
taken with considerable caution. As we will see below, surface re-relaxations,
not only of the nearest neighbor atoms of the adsorption site but also
including atoms in the next surface unit cells, make up a major contribution
to the binding energy of the molecules, in particular for the dissociated
state. Due to the  limited cluster size it is questionable how well such
structural effects have been captured in this study since the hydroxylated
bridging O atoms created by the dissociation of a water molecule were already
located at the boundary of the cluster.\cite{ThW5}

The main intention of our study is not to resolve the controversy between
the different DFT studies on the molecular or dissociative nature of the
adsorption of water, but to evaluate the relative stability of surface
hydroxyl groups created by the non-reductive dissociation of water compared
to both molecular water adsorption and to hydroxylation of the surface by
reductive hydrogen adsorption which we will represent in terms of a phase
diagram. The general features of the phase diagram and the qualitative
position of the phase boundaries only depend on the overall interaction
strength of the water molecules with the TiO$_2$(110) surface but not,
in cases where the energy difference between molecular and dissociative
adsorption is small as for single, isolated water molecules, whether the
molecular or the dissociate state turns out to be more stable. However,
since our calculational setup with a large surface unit cell and a passivated
slab seems to be quite reliable to give well converged results, we explored
the question whether water dissociates on the ideal terraces of the
TiO$_2$(110) surface or not a little bit further than it would have been
necessary for the construction of our surface phase diagram.

The most favorable adsorption site for water molecules and hydroxyl groups
is ontop of the fivefold-coordinated Ti(5) cations, thereby restoring their
sixfold-coordination as in the bulk. Water molecules lie relatively flat
on the surface and form a weak hydrogen bond to one of the bridging O(2)
atoms. Hydroxyl groups take a more upright position and orient themselves 
along a Ti(5)--O(3) bond, as long as they cannot form a hydrogen bond
to an adsorbed molecule on a neighboring Ti(5) site (see Fig.~\ref{F9}).

\begin{figure}[!t]
\noindent
\includegraphics[width=246pt]{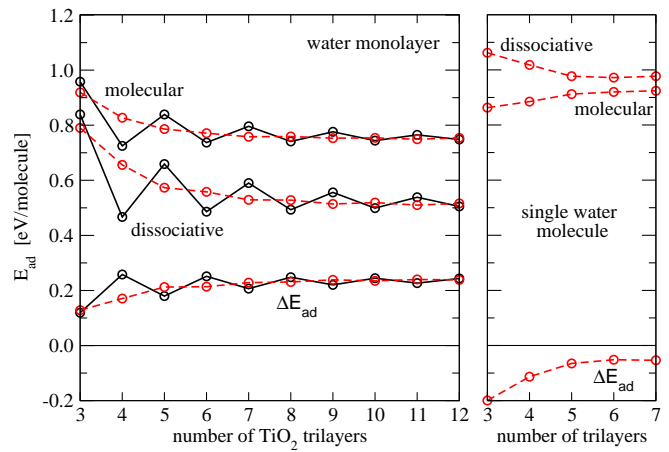}
\caption{\label{F10}
Convergence of the molecular and dissociative water adsorption energy
$E_{\rm ad}$ and the difference $\Delta E_{\rm ad}$=$E_{\rm ad}^{\rm mol}$
$-$$E_{\rm ad}^{\rm diss}$ with slab thickness at full monolayer coverage
(left) and for single water molecules in a (4$\times$2) surface unit cell
(right). The solid black lines represent results from full relaxations with
water adsorbed symmetrically on both sides of the slab. The calculations in
which water was adsorbed only on one side of the slab with the bottom two
trilayers fixed at the bulk positions and the broken surface bonds at the
bottom of the slab saturated with pseudo-hydrogen atoms are shown by the
red dashed lines.}
%Adsorption energy of H$_2$O for dissociative (solid black line) and molecular
%(dashed red line) adsorption as function of the water coverage of the surface.
\end{figure}

For single water molecules in a (4$\times$2) surface unit cell we find
that dissociative adsorption is preferred compared to the molecular state
by 0.11\,eV, as can be seen from the compiled results in Table~\ref{T6}.
This is in contrast to the results of Ref.~\onlinecite{ExW2,ThW9,TV05},
but in agreement with Ref.~\onlinecite{ThW8,ExW3,ExW4}. On the other
hand, at full monolayer coverage molecular adsorption is found to be
0.05\,eV and 0.19\,eV per water molecule more stable than partial or full
dissociation, respectively. In view of the very contradictory results from
previous DFT calculations in the literature on the relative stability of
molecular and dissociative adsorption of water on TiO$_2$(110), we have
carefully re-evaluated the convergence of our computational setup. From the
surveil of the literature it appears that the slab thickness and the degree of
atomic relaxation of the slabs are the most crucial parameters. Therefore we
have recalculated the energy difference $\Delta E_{\rm ad}$ between molecular
and dissociative adsorption of water as a function of slab thickness. In the
first set of calculations we have considered a full monolayer water coverage
as represented by one water molecule in a (1$\times$1) surface unit cell. As
shown in Fig.~\ref{F10}, the molecular and dissociative adsorption energy
$E_{\rm ad}^{\rm mol}$ and $E_{\rm ad}^{\rm diss}$, respectively, exhibit the
well-known strong odd--even oscillations if the water molecules are adsorbed
symmetrically on both sides of the slab and a full atomic relaxation is
performed. These oscillations are even present in the energy difference
$\Delta E_{\rm ad}$=$E_{\rm ad}^{\rm mol}$$-$$E_{\rm ad}^{\rm diss}$.
Switching to our standard setup in which we adsorb water only on one side of
the slab, fix the atoms in the two bottom layers at the bulk positions and
saturate the broken surface bonds at the bottom with pseudo-hydrogen atoms,
the odd--even oscillations are almost completely removed (see Fig.~\ref{F10}).
For $\Delta E_{\rm ad}$ we find only a small increase from 0.19\,eV for the
four trilayer slab to 0.23\,eV in the limit of infinite slab thickness. In
the second set of calculations we repeated the calculations for a single water
molecule in a (4$\times$2) surface unit cell with our passivated slabs for
slab thicknesses up to seven trilayers. In the seven trilayer calculation the
supercell contained 355 atoms. Also in this case, as shown in Fig.~\ref{F10},
there are almost no odd--even oscillations visible in the water adsorption
energy $E_{\rm ad}$ and the energy difference $\Delta E_{\rm ad}$ between the
molecular and dissociative adsorption increases only slightly from $-$0.11\,eV
for four trilayers to $-$0.06\,eV in the six and seven trilayer
calculation. Since in general such small energy differences depend sensitively
on the quality of the pseudopotentials and the functional which are used in
the computations, we only conclude that molecular and dissociative adsorption
become energetically degenerate in the limit of single, isolated water molecules.

When comparing our results on the adsorption of isolated water molecules
with the literature surveil in Table~\ref{T5} it has to be taken into account
that in the previous calculations for `single' water molecules smaller surface
unit cells with effective water coverages of 1/4 monolayer were used, compared
to 1/8 monolayer in our study. As can be seen from our results in
Table~\ref{T6}, at 1/4 monolayer coverage there is already a significant
interaction between the water molecules. For two water molecules in the
(4$\times$2) surface unit cell the water binding energy is clearly reduced
compared to a single molecule. The reason is, as we will point out below, the
large contribution of the surface relaxations to the binding energy, which is
now smaller since less degrees of freedom are associated with each molecule.
In particular, for a (2$\times$2) arrangement of water molecules, which was
used in basically all previous calculations as setup for `isolated' molecules,
we find already a preference for molecular adsorption. Clearly, the 1/4
monolayer coverage does not represent isolated water molecules, and in
principle, it would have to be checked if our 1/8 monolayer truly represents
the limit of independent molecules. This implies that also in the embedded
cluster approach large clusters are needed to take these contributions from
the atomic relaxations appropriately into account.

Our result that for isolated water molecules molecular and dissociative
adsorption is energetically degenerate is not in agreement with the common
view expressed in most experimental studies that water only dissociates at
defect sites. Since our calculations are very well converged, this might point
towards a deficiency of the pseudopotential approximation or, in general,
PBE/DFT calculations. However, it also has to be taken into account that all
experiments have shown a certain amount of dissociated molecules and it is
very difficult to distinguish whether they only stem from dissociation at
defects or if some initial dissociation at very low coverages has taken place.

Comparing the water binding energies of two water molecules in a c(4$\times$2)
with the p(4$\times$1) arrangement we find that the interaction of water
molecules between different [001] rows is rather weak since the relative
position of the water molecules within the [001] rows does not matter. This
is very different from the situation when both water molecules are placed into
the same [001] row (as in the p(2$\times$2) configuration), where the changes
in the surface relaxation lead to a strong repulsion between dissociated
molecules. This repulsion has also been reported by Lindan {\it et al}.,%
\cite{ThW8}  although with a slightly smaller magnitude. The repulsion can be
compensated by adsorbing the two water molecules on neighboring Ti(5) sites,
since now a hydrogen bond between the molecules is formed. We have considered
all possible pairs of dissociated and undissociated water molecules with
`upward' and `downward' orientation as illustrated in Fig.~\ref{F9}.
As can be inferred from the binding energies in Table~\ref{T6}, all these
configurations are almost equivalent, with a slight preference of pairs of
dissociated/undissociated molecules in which the donating molecule of the
hydrogen bond stays intact and the receiving molecule dissociates.
Furthermore, we can see that the relative orientation of the molecules within
the pair (`upward' or `downward') does not matter, which will be also true at
higher water coverages.

We take now these water pairs as building blocks to form higher water
coverages. First we put two water pairs in the same [001] row to built
structures of alternating full and empty [001] lines. The molecular
adsorption is now the most stable one. The half-dissociated structure
of Lindan {\it et al.}\cite{ThW4} is only slightly lower in energy.
The fully dissociated structure, however, has become clearly unfavorable.
Since the interaction of the water molecules between [001] rows is weak,
as already pointed out above, this result remains unchanged when all
[001] rows are filled with water molecules and the full monolayer
coverage is reached.

For the length of the Ti(5)--O$_{\rm water}$ bond we obtained values between
2.246\,{\AA} and 2.306\,{\AA} for isolated water molecules and at full
monolayer coverage, respectively. In a recent quantitative analysis of
photoelectron diffraction experiments\cite{AB05} a value for this bond length
of 2.21$\pm$0.02\,{\AA} was determined. Taking into account that GGA typically
overestimates bond length by 1--2\%, our result for isolated water molecules
agrees quite well with the experimental value, but overestimate it slightly in
the limit of full monolayer coverage.

In conclusion, we find that in the limit of very low coverages the molecular
and dissociative adsorption of water is almost degenerate. For single,
isolated water molecules we find a small preference for dissociation, however,
already at coverages when water pairs can form, mixed dissociated/molecular
structures are more favorable than full dissociation. Finally, in the full
monolayer coverage limit the molecular adsorption is the most stable
configuration. Based on these results  one would expect that if water islands
are formed upon adsorption, almost all water molecules stay intact and only
those molecules at the boundary of the water patch which do not have a
neighbor to which they can form a hydrogen bond may dissociate. It is
interesting to see the difference in the role of donated and accepted hydrogen
bonds for the stability of water molecules between the TiO$_2$(110) and
ZnO(10$\bar{1}$0) surface. On ZnO(10$\bar{1}$0), water molecules also show a 
tendency to dissociate.\cite{ZnW1,ZnW2,ZnW3}  But while on ZnO(10$\bar{1}$0)
isolated water molecules stay intact and only start to dissociate when they
receive an hydrogen bond, isolated water molecules on TiO$_2$(110) are found
to be close to dissociation and recombination is preferential when they can
donate a hydrogen bond. 

Our results for the water binding energies are throughout slightly lower than
the previous results obtained with the PW91 functional,\cite{ThW3,ThW4,ThW9,ThWA}
but significantly higher than what has been found in RPBE calculations%
\cite{ExW2,ExW3,ExW4,ThW7,ThW8,TV05} (see also Tables~\ref{T4} and \ref{T5}).
In order to compare our water binding energy of 0.82\,eV at full monolayer
coverage with experiment, we have furthermore calculated the corrections due
to the zero-point vibration energy (ZPE). The vibration frequencies of the
surface with and without the molecular water monolayer were obtained in
harmonic approximation by a finite difference scheme. Only the atoms in the
adsorbate and the first surface layer were displaced by 0.01\,{\AA} in the
three cartesian directions. For the ZPE we obtain 0.12\,eV so that we arrive
at a ZPE-corrected water binding energy at monolayer coverage of 0.70\,eV,
which is in excellent agreement with the results of the TDS experiments of
0.65\,eV\cite{ExW6} and 0.66\,eV\cite{ExW7}. The minor overbinding of the
water molecules is typical for PBE calculations and has also been observed for
the adsorption of water on ZnO(10$\bar{1}$0), where PBE calculations gave a
water binding energy of 1.13\,eV (without ZPE corrections) compared to the 
TDS/Redhead analysis of 1.03\,eV.\cite{ZnW1} 

Looking at the qualitative trend of the adsorption energy with water coverage
of the surface it is interesting to see in Table~\ref{T6} that overall the
binding energy decreases with water saturation. This contradicts the results
of Ref.~\onlinecite{ThW9},  whereas it is in full agreement with the RPBE
calculations of Lindan {\it et al}.\cite{ThW7,ThW8}  and, in addition, to the
experimental observation in the TDS measurements. This is a rather unusual 
behavior since one would expect an increase in the binding energy with water
coverage due to the formation of hydrogen bonds. In order to gain deeper
insights into the origin of this behavior we have decomposed the process of
desorbing the water molecules into four steps. Starting from the fully relaxed
adsorbate structure we detach in the first step the water layer from the
surface without relaxing the atoms. Then we separate the water molecules in
order to obtain the hydrogen bond strength. Finally, we allow the TiO$_2$(110)
surface and the water molecules to relax to obtain the relaxation energies of
the systems (see Ref.~\cite{ZnW3} for a similar analysis of the water
adsorption energy on ZnO(10$\bar{1}$0) ). For structures with neighboring
water molecules we find indeed that hydrogen bonds are formed which increases
the water binding energy. This gain in adsorption energy, however, is
overcompensated at higher coverages by a loss in relaxation energy of the
substrate per adsorbate molecule. Why are the surface relaxations so much
more important for the TiO$_2$(110) surface than, for example,
ZnO(10$\bar{1}$0)? As we have seen in Sec.~\ref{method}, the surface
relaxation energy of the stoichiometric, ideal surface is quite large,
1.21\,eV per surface unit cell compared to 0.37\,eV for ZnO(10$\bar{1}$0).
Thus, re-relaxations induced by the restoration of the cation bulk
coordination upon water adsorption will have a much stronger impact on
the binding energy of the molecules than for ZnO(10$\bar{1}$0).

Lindan {\it et al}.\cite{ThW8}  have suggested that the dissociation of
water at low coverages has not been observed experimentally because the
dissociation of the molecules is hindered by a significant dissociation
barrier which is difficult to overcome on the time scale of the experiments.
Using NEB with a subsequent refinement of the transition state with the
dimer method we have calculated the activation barrier for water dissociation
for two cases: an isolated water molecule in a (4$\times$2) surface cell
and a single water molecule within a molecularly adsorbed water layer (also
employing a (4$\times$2) supercell). Our results of 0.16\,eV and 0.14\,eV,
respectively, are significant lower than the activation barriers reported
by Lindan {\it et al}.\cite{ThW8} of 0.45\,eV (single water molecule in
a (4$\times$1) cell) and 0.26\,eV (dissociation of every second water molecule
in a full monolayer) and show a much weaker dependence on the water coverage.
Therefore, based on our results, we cannot confirm that the dissociation of
water molecules at low coverages is prevented by a large dissociation barrier.

%------------------------------------------------------------------------------

\subsection{Surface Phase Diagram}

\begin{figure}[!t]
\noindent
\includegraphics[width=246pt]{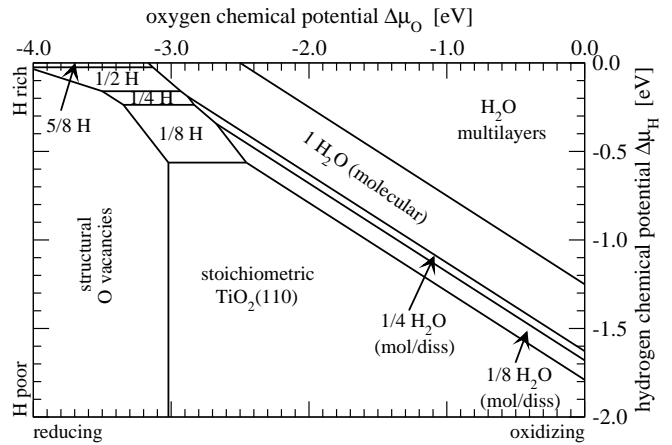}
\caption{\label{F11}
Phase diagram for the TiO$_2$(110) surface in thermodynamic equilibrium with
H$_2$ and O$_2$ particle reservoirs controlling the chemical potentials
$\Delta\mu_{\rm H}$  and $\Delta\mu_{\rm O}$.}
\end{figure}

Finally we combine the results of all previous sections to construct a
two-dimensional phase diagram. We assume that the TiO$_2$(110) surface is
simultaneously in thermodynamic equilibrium with reservoirs with which it can
exchange O and H atoms, for example, a surrounding O$_2$ and H$_2$ gas phase.
The surface free energy $\Delta\gamma$ of the different TiO$_2$(110)
surface structures now depends on both chemical potentials,
$\Delta\mu_{\rm O}$ and $\Delta\mu_{\rm H}$.  By indicating the most stable
structure and composition of the TiO$_2$O(110) surface as a function of
the two chemical potentials we obtain the phase diagram which is shown
in Fig.~\ref{F11}.

The surface phase diagram of Fig.~\ref{F11} summarizes in a condensed
fashion the results of our study on the relative stability of the different
surface compositions which we have considered. The phase diagram is dominated
by four different types of surface structures: the stoichiometric, ideal
surface, the water saturated, oxidized surface and the reduced surface
structures by O depletion and H adsorption. For ambient conditions
($\Delta\mu_{\rm O}=-1.1$\,eV) we find that the TiO$_2$(110) surface is
saturated with molecularly adsorbed water. The water can be gradually removed
by lowering the O and H chemical potentials (i.e., by heating or lowering the
partial pressures). Passing through a transition region with low water
coverages in which dissociated water molecules may appear, the stoichiometric,
ideal surface structure is reached. It is the most stable surface structure at
UHV conditions for a wide range of temperatures ($\Delta\mu_{\rm O}=-1.9$\,eV,
$\Delta\mu_{\rm H}=-1.5$\,eV at room temperature). By heating the surface in
UHV, the surface can be reduced by removing bridging oxygen atoms and creating
structural O~vacancies. At hydrogen rich and oxygen poor conditions the
surface is reduced by hydrogen adsorption. However, the surface cannot be
fully hydroxylated, but only a maximum H coverage of about 60\% can be
reached.

Overall the surface phase diagram shows a quite expected behavior of the
TiO$_2$(110) surface and does not yield any new surprises. Under UHV
conditions the stoichiometric and the O reduced surface structures are
the most important ones. On the other hand, under wet conditions or at high
hydrogen partial pressures, as it is typical in catalytic or photochemical
applications, the surface is water covered or reduced via hydroxylation.

%------------------------------------------------------------------------------

\section{Summary and Conclusions}

Using density functional theory we have investigated in detail the formation
of oxygen-related vacancies as well as the interaction of hydrogen and water
with the TiO$_2$(110) surface in terms of a comprehensive surface phase
diagram.
Careful convergence tests have shown that a computational setup of using slabs
consisting of four trilayers in conjunction with saturating the surface bonds
at the bottom with artificial hydrogen-like atoms with effective nuclear
charges of +4/3 and +2/3 and fixing the atoms in the bottom two trilayers
at their bulk positions almost completely eliminates the otherwise pronounced 
odd-even oscillations of surface and adsorption energies as a function of 
slab thickness.
Using this procedure, the relaxation energy of the stoichiometric surface
as well as the H and H$_2$O adsorption energies at full monolayer coverage
are within 0.06\,eV per surface unit cell, 0.02\,eV per H atom and 0.04\,eV
per H$_2$O molecules, respectively, of the extrapolated results for infinitely
thick slabs.
Thus, very well converged results with slab thickness are obtained without
relying on somewhat questionable extrapolation schemes such as the ``25\%
rule''\cite{ThW9} or averaging four and five trilayer results,\cite{Ovac8}
as proposed previously in the literature.

It is also demonstrated that the contribution of surface relaxations to 
both binding and formation energies of molecules and defects, respectively,
are substantial.
Both adsorbate- and defect-induced relaxations are not limited to one surface
unit cell, but contributions to the relaxation energy from more neighboring
surface cells have to be taken into account. This leads to substrate-mediated
interactions between defects and adsorbate molecules which can extend over
several lattice constants. Consequently, large surface unit cells have to be
used when properties of isolated defects or adsorbed molecules are studied.
The same arguments apply for embedded cluster studies where clusters have to
be large enough to be able to take the relaxations of the relevant environment
appropriately into account.

In general, GGA/DFT calculations of reduced TiO$_2$ structures are hampered
by the underestimation of the TiO$_2$ band gap of about 1.0\,eV. For isolated
O~vacancies and pairs of OH groups on the TiO$_2$(110) surface it has been
shown that in GGA/DFT calculations the defect level lies at the bottom of
the conduction band\cite{GP07,VP06} instead of 0.7--0.9\,eV below. 
Naively one would therefore expect that GGA/DFT calculations should
underestimate the energy of reduced TiO$_2$ structures and thus
{\em overestimate} the reducibility of TiO$_2$ since defect states
will tend to be too delocalized.
In contrast to this expectation, a survey of the literature data showed
instead that GGA/DFT overestimates the O~vacancy formation energy.
Furthermore we found in the present study that the dissociation energy
of water at O~vacancies is slightly too low in the PBE/DFT calculations
and also the predicted maximum coverage of the TiO$_2$(110) surface 
with hydrogen atoms is found to be slightly smaller than observed in
experiment.
This implies that in all three cases the GGA/DFT calculation have
{\em underestimated} the reducibility of the TiO$_2$(110) surface. 
However, the presumable error in the total energies is small: about 0.3\,eV
for the O defect formation energy and a tenth of an eV in the other two cases. 
Based on this analysis we conclude that the {\em energetics} is quite reliably
described by GGA/DFT despite the problem of such calculations to obtain
accurately the underlying electronic structure of the corresponding reduced
TiO$_2$ surfaces.
It might be argued that the energetics of reduced TiO$_2$ is well described
because although the defect level of O~vacancies and OH pairs is at the bottom
of the conduction band instead of 0.7--0.9\,eV below, its energy difference
with respect to the top of the valence band is almost right. This might be
correct for single, isolated O~vacancies and OH pairs. However, in the case
of an increasing hydroxylation of the surface when the defect states start
to interact and to hybridize with the conduction band the whole defect band
is now more easily accessible due to the underestimation of the band gap,
but still the overall energetics is quite reliable with a tendency to
{\em underestimate} the reducibility of TiO$_2$. 

After having clarified these crucial convergence and accuracy issues, 
we have compared the relative stability of surfaces with oxygen vacancies or
adsorbed hydrogen atoms and water molecules by combining our DFT calculations
with a thermodynamic formalism. Lowest-energy structures were determined by
assuming that the surfaces are in thermodynamic equilibrium with oxygen and
hydrogen reservoirs. The exchange of O and H atoms with the reservoirs is
described by introducing appropriate chemical potentials which can be related
to experimental temperature and pressure conditions. This allows us to extend
the zero-temperature and zero-pressure DFT calculations to more realistic
environmental conditions such as in UHV surface science experiments or in
catalytic and photochemical applications.

The central result of this approach is the phase diagram depicted in
Fig.~\ref{F11}. For ambient conditions we find that the non-reductive
adsorption of water prevails. A full monolayer of molecular adsorbed water
is predicted as the most stable adsorbate structure in agreement with
experimental observations.
In the low coverage limit molecular and dissociative adsorption becomes
energetically degenerate. 
In addition, according to our calculations the dissociation barrier for
molecularly adsorbed water molecules is not large enough to prevent their
dissociation even at low temperatures.

At strongly reducing, oxygen-poor conditions structural O~vacancies are formed
on the TiO$_2$(110) surface in thermodynamic equilibrium. However, for a wide
range of temperature and pressure conditions it is found that removing full
TiO$_2$ units is thermodynamically more favorable instead. 
In the case of reduction of the surface via hydrogen adsorption we observed 
that the surface cannot be fully hydroxylated in thermodynamic equilibrium. 
A maximum coverage of about 60\% is predicted from our thermodynamical
analysis. Furthermore, our calculated energy barriers for hydrogen migration
indicate than hydrogen atoms rather migrate into the bulk than desorbing from
the surface.
Both observations are supported by recent experiments in which a maximum
saturation of the surface with hydrogen of 70\% is reported, whereas no
desorption of H$_2$ could be observed in TDS.\cite{YC08}

\begin{acknowledgments}
The authors thank Christof W\"oll, Monica Calatayud, and Ulrike Diebold for
fruitful discussions. This work has been supported by the German Research
Foundation (DFG) via the Collaborative Research Center SFB 558
``Metal-Substrate Interactions in Heterogeneous Catalysis''.
Computational resources were provided by {\sc Bovilab@RUB} (Bochum).
\end{acknowledgments}

%------------------------------------------------------------------------------

%------------------------------------------------------------------------------

\begin{thebibliography}{99}
\bibitem{DB03} U. Diebold, Surf. Sci. Rep. {\bf 48}, 53 (2003).
\bibitem{DV94} M. Ramamoorthy, D. Vanderbilt, and R.D. King-Smith,
               Phys. Rev. B {\bf 49}, 16721 (1994).
\bibitem{TA78} S.J. Tauster, S.C. Fung, and R.L. Garten,
               J. Am. Chem. Soc. {\bf 100}, 170 (1978).
\bibitem{TA87} S.J. Tauster, Acc. Chem. Res. {\bf 20}, 389 (1987).
\bibitem{VS84} M.A. Vannice and C. Sudhakar, J. Phys. Chem. {\bf 88},
               2429 (1984).
\bibitem{ExW1} I.M. Brookes, C.A. Muryn, and G. Thornton,
               Phys. Rev. Lett. {\bf 87}, 266103 (2001).
\bibitem{ExW2} R. Schaub, P. Thostrup, N. Lopez, E. L{\ae}gsgaard,
               I. Stensgaard, J.K. N{\o}rskov, and F. Besenbacher,
               Phys. Rev. Lett. {\bf 87}, 266104 (2001).
\bibitem{ExW3} S. Wendt, R. Schaub, J. Matthiesen, E.K. Vestergaard,
               E. Wahlstr\"om, M.D. Rasmussen, P. Thorstrup, L.M. Molina,
               E. L{\ae}gsgaard, I. Stensgaard, B. Hammer, and
               F. Besenbacher, Surf. Sci. {\bf 598}, 226 (2005).
\bibitem{ExW4} S. Wendt, J. Matthiesen, R. Schaub, E.K. Vestergaard,
               E. L{\ae}gsgaard, F. Besenbacher, and B. Hammer,
               Phys. Rev. Lett. {\bf 96}, 066107 (2006).
\bibitem{ExW5} O. Bikondoa, C.L. Pang, R. Ithnin, C.A. Muryn, H. Onishi,
               and G. Thornton, Nature Mat. {\bf 5}, 189 (2006) 189.
\bibitem{ExW6} M.B. Hugenschmidt, L. Gamble, and C.T. Campbell,
               Surf. Sci. {\bf 302}, 329 (1994).
\bibitem{ExW7} M.A. Henderson, Surf. Sci. {\bf 355}, 151 (1996).
\bibitem{ExW8} D. Brinkley, M. Dietrich, T. Engel, P. Farrall, G. Gantner,
               A. Schafer, and A. Szuchmacher, Surf. Sci. {\bf 395}, 292
               (1998).
\bibitem{ExW9} R.L. Kurtz, R. Stockbauer, T.E. Madey, E. Roman, and
               J.L. De Segovia, Surf. Sci. {\bf 218}, 178 (1989).

\bibitem{DB92} J.-M. Pan, B.L. Maschhoff, U. Diebold, and T.E. Madey,
               J. Vac. Sci. Technol. A {\bf 10}, 2470 (1992).
\bibitem{DB05} M. Batzill, K. Katsiev, J.M. Burst, U. Diebold, A.M. Chaka,
               B. Delley, Phys. Rev. B {\bf 72}, 165414 (2005).
\bibitem{BM07} K. Katsiev, M. Batzill, U. Diebold, A. Urban, and B. Meyer,
               Phys. Rev. Lett. {\bf 98}, 186102 (2007).
\bibitem{HK81} V.E. Henrich and R.L. Kurtz, Phys. Rev. B {\bf 23}, 6280
               (1981).
\bibitem{KB04} M. Kunat, U. Burghaus, and Ch.\ W\"oll,
               Phys. Chem. Chem. Phys. {\bf 6}, 4203 (2004).
\bibitem{SF00} S. Suzuki, K.I. Fukui, H. Onishi, and Y. Iwasawa,
               Phys. Rev. Lett. {\bf 84}, 2156 (2000).
\bibitem{FK01} T. Fujino, M. Katayama, K. Inudzuka, T. Okuno, and K. Oura,
               Appl. Phys. Lett. {\bf 79}, 2716 (2001).
\bibitem{YC08} X.-L. Yin, M. Calatayud, H. Qiu, Y. Wang, A. Birkner, C. Minot,
               and Ch.\ W\"oll, ChemPhysChem {\bf 9}, 253 (2008).
\bibitem{Ovac0} M. Ramamoorthy, R.D. King-Smith, and D. Vanderbilt,
                Phys. Rev. B {\bf 49}, 7709 (1994).
\bibitem{Ovac1} P.J.D. Lindan, N.M. Harrison, M.J. Gillan, and J.A. White,
                Phys. Rev. B {\bf 55}, 15919 (1997).
\bibitem{Ovac2} A.T. Paxton and L. Thien-Nga, Phys. Rev. B {\bf 57}, 1579
                (1998).
\bibitem{Ovac3} T. Bredow and G. Pacchioni, Chem. Phys. Lett. {\bf 355}, 417
                (2002).
\bibitem{Ovac4} A. Vijay, G. Mills, and H. Metiu, J. Chem. Phys. {\bf 118},
                6536 (2003).
\bibitem{Ovac5} X. Wu, A. Selloni, M. Lazzeri, and S.K. Nayak, Phys. Rev. B
                {\bf 68}, 241402(R) (2003).
\bibitem{Ovac6} M.D. Rasmussen, L.M. Molina, and B. Hammer, J. Chem. Phys.
                {\bf 120}, 988 (2004).
\bibitem{Ovac7} X. Wu, A. Selloni, and S.K. Nayak, J. Chem. Phys. {\bf 120},
                4512 (2004).
\bibitem{Ovac8} J. Oviedo, M.A. San Miguel, and J.F. Sanz, J. Chem. Phys.
                {\bf 121}, 7427 (2004).
\bibitem{Ovac9} S.-G. Wang, X.-D. Wen, D.-B. Cao, Y.-W. Li, J. Wang, and
                H. Jiao, Surf. Sci. {\bf 577}, 69 (2005).
\bibitem{OvacA} Y.-F. Zhang, W. Lin, Y. Li, K.-N. Ding, and J.-Q. Li,
                J. Phys. Chem. B {\bf 109}, 19270 (2005).
\bibitem{OvacB} K. Hameeuw, G. Cantele, D. Ninno, F. Trani, and G. Iadonisi,
                phys. stat. sol. (a) {\bf 203}, 2219 (2006).
\bibitem{OvacC} T. Pabisiak and A. Kiejna, Solid State Comm. {\bf 144}, 324
                (2007).
\bibitem{GP07} M.V. Ganduglia-Pirovano, A. Hofmann, and J. Sauer,
               Surf. Sci. Rep. {\bf 62}, 219 (2007)
\bibitem{LM02} J. Leconte, A. Markovits, M.K. Skalli, C. Minot, and
               A. Belmajdoub, Surf. Sci. {\bf 497}, 194 (2002).
\bibitem{ThW1} J. Goniakowski and M.J. Gillan, Surf. Sci. {\bf 350}, 145
               (1996).
\bibitem{ThW2} P.J.D. Lindan, N.M. Harrison, J.M. Holender, and M.J. Gillan,
               Chem. Phys. Lett. {\bf 261}, 246 (1996).
\bibitem{ThW3} S.P. Bates, G. Kresse, and M.J. Gillan, Surf. Sci. {\bf 409},
               366 (1998).
\bibitem{ThW4} P.J.D. Lindan, N.M. Harrison, and M.J. Gillan,
               Phys. Rev. Lett. {\bf 80}, 762 (1998).
\bibitem{ThW5} E.V. Stefanovich and T.N. Truong, Chem. Phys. Lett. {\bf 299},
               623 (1999).
\bibitem{ThW6} W. Langel, Surf. Sci. {\bf 496}, 141 (2002).
\bibitem{ThW7} C. Zhang and P.J.D. Lindan, J. Chem. Phys. {\bf 118}, 4620
               (2003).
\bibitem{ThW8} P.J.D. Lindan and C. Zhang, Phys. Rev. B {\bf 72}, 075439
               (2005).
\bibitem{ThW9} L.A. Harris and A.A. Quong, Phys. Rev. Lett. {\bf 93}, 086105
               (2004); Comment: P.J.D. Lindan and C. Zhang, Phys. Rev. Lett.
               {\bf 95}, 029601 (2005); Reply: L.A. Harris, and A.A. Quong,
               Phys. Rev. Lett. {\bf 95}, 029602 (2005).
\bibitem{ThWA} H. Perron, J. Vandenborre, C. Domain, R. Drot, J. Roques,
               E. Simoni, J.-J. Ehrhardt, and H. Catalette, Surf. Sci.
               {\bf 601}, 518 (2007).
\bibitem{KP87} E. Kaxiras, Y. Bar-Yam, J.D. Joannopoulos, and K.C. Pandey,
               Phys. Rev. B {\bf 35}, 9625 (1987).
\bibitem{QM88} G.-X. Qian, R.M. Martin and D.J. Chadi, Phys. Rev. B {\bf 38},
               7649 (1988).
\bibitem{RS01} K. Reuter and M. Scheffler, Phys. Rev. B {\bf 65}, 035406
               (2001).
\bibitem{BM04} B. Meyer, Phys. Rev. B  {\bf 69}, 045416 (2004).
\bibitem{CP1} R. Car and M. Parrinello, Phys. Rev. Lett. {\bf 55}, 2471 (1985).
\bibitem{CP2} D. Marx and J. Hutter, in: {\it Modern Methods and Algorithms of
              Quantum Chemistry}, edited by J. Grotendorst, NIC, FZ J\"ulich,
              p.~301 (2000);\\ see: {\tt www.theochem.rub.de/go/cprev.html};\\
              CPMD code: J. Hutter {\it et al.}, see: {\tt www.cpmd.org}.
\bibitem{PBE} J.P. Perdew, K. Burke, and M. Ernzerhof, Phys. Rev. Lett.
              {\bf 77}, 3865 (1996); Erratum: Phys. Rev. Lett. {\bf 78},
              1396 (1997).
\bibitem{DV90} D. Vanderbilt, Phys. Rev. B {\bf 41}, 7892 (1990).
\bibitem{BG04} T. Bredow, L. Giordano, F. Cinquini, and G. Pacchioni,
               Phys. Rev. B {\bf 70}, 035419 (2004).
\bibitem{TL06} S.J. Thompson and S.P. Lewis, Phys. Rev. B {\bf 73}, 073403
               (2006).
\bibitem{HC06} K.J. Hameeuw, G. Cantele, D. Ninno, F. Trani, and G. Iadonisi,
               J. Chem. Phys. {\bf 124}, 024708 (2006).
\bibitem{KP07} A. Kiejna, T. Pabisiak, and S.W. Gao, J. Phys.: Condens.
               Matter {\bf 18}, 4207 (2007).
\bibitem{AB71} S.C. Abrahams and J.L. Bernstein, J. Chem. Phys. {\bf 55},
               3206 (1971).
\bibitem{BA00} I.G. Batyrev, A. Alavi, and M.W. Finnis, Phys. Rev. B {\bf 62},
               4698 (2000).
\bibitem{Com1} This deviation is due to the O pseudopotential. By decreasing
               the cut-off radius, making the local part of the potential more
               attractive and adding $d$-projectors, the converged
               all-electron result for the O$_2$ binding energy can be
               systematically approached. However, such pseudopotentials are
               usually much harder and require a larger plane wave cut-off
               energy.
\bibitem{NIST} {\it NIST Chemistry WebBook}, edited by P.J. Linstrom and
               W.G. Mallard, NIST Standard Reference Database No.~69
               (National Institute of Standards and Technology, Gaithersburg
               MD, 2001), see: http://webbook.nist.gov
\bibitem{P08} G. Pacchioni, J. Chem. Phys. {\bf 128}, 182505 (2008).
\bibitem{GC92} K.M. Glassford and J.R. Chelikowsky, Phys. Rev. B {\bf 46},
               1284 (1992).
\bibitem{PC77} J. Pascual, J. Camassel, and H. Mathieu, Phys. Rev. Lett.
               {\bf 39}, 1490 (1977); Phys. Rev. B {\bf 18}, 5606 (1978).
\bibitem{GA84} W. G\"opel, J.A. Anderson, D. Frankel, M. Jaehnig, K. Phillips,
               J.A. Sch\"afer, and G. Rocker, Surf. Sci. {\bf 139}, 333 (1984).
\bibitem{H98} M.A. Henderson, Surf. Sci. {\bf 400}, 203 (1998).
\bibitem{VP06} C. Di Valentin, G. Pacchioni, and A. Selloni, Phys. Rev. Lett.
                {\bf 97}, 166803 (2006).
\bibitem{HE03} M.A. Henderson, W.S. Epling, C.H.F. Peden, and C.L. Perkins,
               J. Phys. Chem. B {\bf 107}, 534 (2003).
\bibitem{LF91} M. Lannoo and P. Friedel, {\it Atomic and Electronic Structure
               of Surfaces}, Springer Verlag, Berlin (1991).
\bibitem{MD04} B. Meyer and D. Marx, Phys. Rev. B {\bf 69}, 235420 (2004).
\bibitem{Com2} We follow here the notation of Ref.~[\onlinecite{DB03}]: a
               ($n$$\times$$m$) cell is built by repeating the primitive
               surface unit cell $n$ and $m$ times along the bulk [001] and
               [1$\bar{1}$0] directions, respectively. It has dimensions of
               $nc \times m\sqrt{2}a$. In some other publications the order
               of $n$ and $m$ is reversed.
\bibitem{BM03} B. Meyer and D. Marx, Phys. Rev. B {\bf 67}, 035403 (2003).
\bibitem{JM98} H. J\'onsson, G. Mills, and K.W. Jacobsen, in: {\it Classical
               and Quantum Dynamics in Condensed Phase Simulations},
               edited by B.J. Berne, G. Ciccotti, and D.F. Coker, World
               Scientific, Singapore, p.~385 (1998).
\bibitem{HJ99} G. Henkelman and H. J\'onsson, J. Chem. Phys. {\bf 111}, 7010
               (1999).
\bibitem{HJ0a} G. Henkelman, B.P. Uberuaga, and H. J\'onsson, J. Chem. Phys.
               {\bf 113}, 9901 (2000).
\bibitem{HJ0b} G. Henkelman and H. J\'onsson, J. Chem. Phys. {\bf 113}, 9978
               (2000).
\bibitem{HB05} A. Heyden, A.T. Bell, and F.J. Keil, J. Chem. Phys. {\bf 123},
               224101 (2005).
\bibitem{KM07} R. Kov\'a\v{c}ik, B. Meyer, and Dominik Marx, Angew. Chem.
               {\bf 119}, 4980 (2007); Angew. Chem. Int. Ed. {\bf 46}, 4894
               (2007).
\bibitem{WB08} S. Wendt, P.T. Sprunger, E. Lira, G.K.H. Madsen, Z. Li,
               J.{\O}. Hansen, J. Matthiesen, A. Blekinge-Rasmussen,
               E. L{\ae}gsgaard, B. Hammer, and F. Besenbacher, Science
               {\bf 320}, 1755 (2008).
\bibitem{RH62} P.A. Redhead, Vacuum {\bf 12}, 203 (1962).
\bibitem{PG01} K. R. Paserba and A. J. Gellman, Phys. Rev. Lett. {\bf 86},
               4338 (2001).
\bibitem{B88} A.D. Becke, Phys. Rev. A {\bf 38}, 3098 (1988).
\bibitem{P86} J.P. Perdew, Phys. Rev. B {\bf 33}, 8822 (1986); Erratum:
              Phys. Rev. B {\bf 34}, 7406 (1986).
\bibitem{PW91} J.P. Perdew, J.A. Chevary, S.H. Vosko, K.A. Jackson,
               M.R. Pederson, D.J. Singh, and C. Fiolhais, Phys. Rev. B
               {\bf 46}, 6671 (1992); Erratum: Phys. Rev. B {\bf 48}, 4978
               (1993).
\bibitem{RPBE} B. Hammer, L.B. Hansen, and J.K. Norskov, Phys. Rev. B
               {\bf 59}, 7413 (1999).
\bibitem{TV05} A. Tilocca, C. Di Valentin, and A. Selloni, J. Phys. Chem. B
               {\bf 109}, 20963 (2005).
\bibitem{Com3} In the calculation of the transition path for water
               dissociation at O~vacancies we have not pre-adsorbed the
               water molecules in the defect and then searched for a
               subsequent dissociation barrier as it was done in
               Ref.~\onlinecite{KMK07} but have started from a water
               molecule in the gas phase.
\bibitem{KMK07} S. Kajita, T. Minato, H.S. Kato, M. Kawai, and T. Nakayama,
               J. Chem. Phys. {\bf 127}, 104709 (2007).
\bibitem{HD03} G.S. Herman, Z. Dohn\'alek, N. Ruzycki, and U. Diebold,
               J. Phys. Chem. B {\bf 107}, 2788 (2003).
\bibitem{AB05} F. Allegretti, S. O'Brien, M. Polcik, D.I. Sayago, and
               D.P. Woodruff, Phys. Rev. Lett. {\bf 95} 226104 (2005).
\bibitem{ZnW1} B. Meyer, D. Marx, O. Dulub, U. Diebold, M. Kunat,
               D. Langenberg, and Ch. W\"oll, Angew. Chem. {\bf 116}, 6809
               (2004); Angew. Chem. Int. Ed. {\bf 43}, 6641 (2004).
\bibitem{ZnW2} O. Dulub, B. Meyer, and U. Diebold, Phys. Rev. Lett. {\bf 95},
               136101 (2005).
\bibitem{ZnW3} B. Meyer, H. Rabaa, and D. Marx, Phys. Chem. Chem. Phys.
               {\bf 8}, 1513 (2006).
%
%\bibitem{Cox} P.A. Cox, {\it Transition Metal Oxides -- An Introduction to
%              Their Electronic Structure}, Clarendon, Oxford (1992).
%\bibitem{SP71} D.R. Stull and H. Prophet, {\it JANAF Thermochemical Tables},
%               Natl. Bur. Stan. (U.S.), 2nd ed. (U.S. GPO Washington, D.C.,
%               1971) 
%
\end{thebibliography}
\end{document}